\documentclass[preprint, authoryear]{elsarticle}
\usepackage{natbib}
\usepackage{amsmath, amssymb, aas_macros}

\begin{document}

\newcommand{\gr}{$\gamma$-ray}
\newcommand{\grs}{$\gamma$-rays}
\newcommand{\hess}{H.E.S.S.}
\newcommand{\pois}{{\rm Pois}}

\title{Handling Systematic Uncertainties and Combined Source Analyses for Atmospheric Cherenkov Telescopes}
\author[a]{Hugh Dickinson\corref{cor1}}
\ead{hugh.dickinson@fysik.su.se}
\author[a]{and Jan Conrad}

\address[a]{Oskar Klein Centre, Department of Physics, Stockholm University,\\
Albanova University Center, SE-10691 Stockholm, Sweden}

\cortext[cor1]{Corresponding author}


\begin{abstract}
In response to an increasing availability of statistically rich observational data sets, the performance and applicability of traditional Atmospheric Cherenkov Telescope analyses in the regime of systematically dominated measurement uncertainties is examined. In particular, the effect of systematic uncertainties affecting the relative normalisation of fiducial ON and OFF-source sampling regions - often denoted as $\alpha$ - is investigated using combined source analysis as a representative example case. The traditional summation of accumulated ON and OFF-source event counts is found to perform sub-optimally in the studied contexts and requires careful calibration to correct for unexpected and potentially misleading statistical behaviour. 
More specifically, failure to recognise and correct for erroneous estimates of $\alpha$ is found to produce substantial overestimates of the combined population significance which worsen with increasing target multiplicity. 
An alternative \emph{joint likelihood} technique is introduced, which is designed to treat systematic uncertainties in a uniform and statistically robust manner. This alternate method is shown to yield dramatically enhanced performance and reliability with respect to the more traditional approach.
\end{abstract} 

\maketitle

\section{Introduction}\label{sec:intro}

Recent developments involving Imaging Atmospheric Cherenkov Telescopes (IACTs) have revolutionised the field of Very High Energy (VHE) \gr\ astronomy. Indeed, observational data obtained using the current generation of stereoscopic arrays have contributed over 100 new sources to the catalogue of confirmed GeV-TeV emitters\footnote{TeVCat: http://tevcat.uchicago.edu}. 

Notwithstanding these impressive achievements, the motivation for astronomers to continue identifying and categorising additional \gr\ sources and source populations remains undiminished. The initial source detections are likely to be dominated by luminous or nearby \gr\ emitters, implying a requirement for subsequent observations to target progressively fainter source populations. Accordingly, the development of analytical techniques which enable maximal exploitation of the available instrumental sensitivity will become increasingly important as the life-cycles of the various IACT arrays unfold. This emerging philosophy is exemplified by the recent work of \citet[][]{2011arXiv1112.0786K} which employs accurate knowledge of the instrumental response to enhance the performance and reliability of an established technique for source detection originally developed by \citet{1983ApJ...272..317L}.

Until recently, the inherent faintness of astrophysical VHE \gr\ fluxes has resulted in statistically limited signal measurements with uncertainties that are dominated by Poissonian shot noise. Accordingly, many traditional IACT analyses have justifiably neglected the existence of previously sub-dominant systematic effects. In contrast, successful source detection at the limits of instrumental sensitivity requires deep observations and the accumulation of rich data sets with abundant event statistics. In this new regime, the relative contribution of statistical fluctuations to the overall error budget is suppressed with measurement accuracy ultimately becoming systematically limited.

This work investigates methods for the appropriate statistical treatment of systematic uncertainties affecting IACT observational data and their subsequent analysis. In Cherenkov astronomy, the synthesis of large data sets typically involves aggregation of temporally sparse event statistics which correspond to distinct values of various observational parameters, each of which can introduce distinct, observation-specific systematic biases.
Accordingly, analyses that are designed to ameliorate the impact of irreducible systematic effects must incorporate sufficient flexibility to successfully model this inter-observational variability.

In the subsequent discussion, the relevant statistical issues and the techniques designed to address them are illustrated using the specific example of \emph{stacking} analysis. In such analyses, measurement data that correspond to multiple, independent but similar experiments are combined to enhance their scientific utility. In the context of \gr\ astronomy, stacking analyses are commonly applied in situations when a target population comprises a large number of individually undetectable source candidates that are theoretically expected to exhibit at least one identical signal characteristic. In the absence of genuine \gr\ emission, the superimposed observational data will be consistent with random Poisson fluctuations about the mean background level. Conversely, multiple faint signals produced by a genuine sample of faint \gr\ sources will reinforce upon combination, yielding a significant overall excess with respect to the measured background. Accordingly, stacking analyses often facilitate the derivation of average source properties for the target population as a whole, even when detection of individual \gr\ signals is rendered impossible by inadequate instrumental sensitivity. The specific consideration of stacking is instructive because combined source analysis inherently segregates subsets of observational data which are likely to incorporate distinct systematic biases. Moreover, stacking analyses provide a more intuitive distinction between systematic effects that undergo temporal variation, such as telescope elevation or optical efficiency, and those that vary on a target-wise basis, such as the ambient night-sky brightness of the number of \gr\ sources within the telescope field-of-view. 

Compelling and theoretically well motivated target populations for combined source analyses include X-ray binary systems \citep{dickinson_thesis}, combined pulsar populations within globular clusters \citep{2009A&A...499..273A} and the self-annihilating dark matter components of dwarf spheroidal galaxies \citep[e.g.][]{2011APh....34..608H}.

\subsection{Relevant Aspects of the Observational Technique}\label{subsec:obs_technique}

The atmosphere of the Earth is almost completely opaque to multi-GeV and TeV radiation and accordingly, the operational mode of IACT arrays relies upon indirect detection of astrophysical \grs. Indeed, the majority of incident VHE photons undergo electron-positron pair production interactions within the Coulomb fields of stratospheric nuclei. Frequently, these interactions initiate electromagnetic cascades of charged particles, which radiate Cherenkov light as they descend superluminally through the dielectric medium of the atmosphere. IACT arrays are designed to image the resultant distribution of Cherenkov photons at ground-level, using the encoded information to reconstruct the properties of the progenitor \gr. The set of reconstructed \gr\ properties define a multi-dimensional parameter space, with suitable sub-spaces that define the event populations that are required for specific astrophysical analyses.

The situation is complicated somewhat by the fact that IACT arrays operate in a strongly background dominated regime and must successfully distinguish a comparatively weak \gr\ signal from an almost overwhelming background of air-showers initiated by hadronic cosmic rays. Fortunately, the properties of hadron- and \gr-initiated air-showers are sufficiently disparate that an effective segregation is possible based on the characteristics of the captured Cherenkov images. Traditionally, a simplified elliptical parameterisation of the Cherenkov light distribution was used in conjunction with lookup tables of simulated air-shower properties to separate the signal and background event populations \citep{1985ICRC....3..453C,hillas,imbackrej}. Modern computational resources permit utilisation of the complete Cherenkov images \cite[e.g][]{1998NIMPA.416..425L,2009APh....32..231D} and the application of advanced machine-learning algorithms \cite[e.g][]{2010APh....34...25F} to further enhance the background rejection efficiency. 

Despite these sophisticated filtering techniques, the sheer number of air-showers that trigger an IACT array during a given observation inevitably results in a subset of hadron-originated Cherenkov images which are sufficiently \gr-like in appearance to survive the event selection procedure. Quantitatively, the probability that a \gr-like event having specific reconstructed parameters will pass background rejection is described by a multi-dimensional \textit{acceptance} function. The reconstructed \gr\ signal ($N_{\rm ON}$) within a suitably defined sampling region of the event property parameter space is a superposition of the true \gr\ signal ($N_{\rm S}$) on an irreducible background ($N_{\rm B}$) of spuriously classified events.
\begin{equation}\label{eq:N_on_N_b}
N_{\rm ON}=N_{\rm S}+N_{\rm B}
\end{equation}
The magnitude of the background component is typically estimated using the number of \gr-like events ($N_{\rm OFF}$) that are reconstructed within one or more nominally OFF-source regions. Adopting this convention, \eqref{eq:N_on_N_b} is normally re-expressed as
\begin{equation}\label{eq:N_on_N_off}
N_{\rm ON}=N_{\rm S}+\alpha N_{\rm OFF}
\end{equation}
where $\alpha$ is a normalisation factor which compensates for any disparities in the instrumental response among the various sampling regions \citep[see e.g.][for a more detailed discussion]{2007A&A...466.1219B}. 

Fundamentally, the precise value of $\alpha$ is dependent upon specific characteristics of the individual observations that contribute to the overall data set. Numerous variable factors such as the target zenith and azimuthal angles, the configuration and functionality of individual telescopes within a larger array, or the ambient atmospheric conditions may modify the nominal system performance. 

The functional dependency of $\alpha$ on the various observational and instrumental parameters is complicated and is must typically be resolved for each observation using a multi-dimensional model estimate of the prevailing system acceptance \citep{2007A&A...466.1219B}. Candidate models may be synthesised by interpolating the distributions of representative \gr-like background data sets, assembled from multiple independent observations of empty fields-of-view. Accurate replication of the true system acceptance using this technique requires an extensive database of suitable observations with characteristics that sample the entire space of observational parameters with adequate resolution. For certain well-understood parameters it may be possible to derive semi-analytic parameterisations of the corresponding systematic distortions, which can improve the interpolation accuracy and reduce the required sampling density. Nonetheless, for complicated instruments with a large number of possible configurations, the observation time required to populate the reference database may be prohibitive. Alternatively, response models which inherently incorporate the appropriate systematic offsets can be constructed on an observation-wise basis using the subset of \gr-like background events that fall outside of the fiducial sampling regions. However, this approach is susceptible to contamination by imperfectly excluded \gr\ sources and is ultimately limited by the availability of event statistics within a single field-of-view. 

For the current generation of IACTs, residual discrepancies between the generated model and the true instrumental response reflect a superposition of imperfectly modelled systematic offsets and are typically at the few percent level \citep{2007A&A...466.1219B}. However, if observational situations arise in which both approaches for modelling the system acceptance are compromised, the resultant disparity could be much larger. For example, if the telescope configuration or the prevailing operating conditions restrict the availability of appropriate archival background data and the target field-of-view is crowded, then systematic effects could dominate statistical uncertainties when searching for faint \gr\ sources. 
Moreover, since the combined impact of multiple systematic effects is often reflected by \textit{non-uniform} variations in the system acceptance, the relative extent and location of the sampling regions within the event property parameter space may also bias the estimated value of $\alpha$. Most experiments tacitly acknowledge the uncertainty in $\alpha$ by attaching a conservative systematic error term to their quoted results. 

\subsection{Alternative Stacking Analyses}\label{subsec:stack_analyses}

Conventional stacking analyses typically entail simple summation of the observed event counts associated with the individual targets in the population sample, with the aggregate datasets forming the basis for subsequent astrophysical inference. However, the background estimation procedures outlined in $\S$\ref{subsec:obs_technique} render this traditional approach (hereafter referred to as \emph{data stacking}) sub-optimal in the context of VHE \gr\ observations. In particular, data stacking analyses are complicated somewhat by the requirement to synthesise a combined  value of the normalisation parameter $\bar{\alpha}$ which corresponds to the stacked dataset. A viable approach uses the definition of the overall signal excess,
\begin{align}\label{eq:excess}
\Delta &= N_{\rm ON} -\bar{\alpha} N_{\rm OFF}\notag\\ &= \displaystyle\sum_{i} N_{{\rm ON},i} - \bar{\alpha}\displaystyle\sum_{i} N_{{\rm OFF},i}, \quad i=1,\ldots, m\notag\\ &= \displaystyle\sum_{i} N_{{\rm ON},i} - \displaystyle\sum_{i} \alpha_{i}N_{{\rm OFF},i}, \quad i=1,\ldots, m
\end{align}
to form an average over all $m$ targets in the sample, weighted by the individual off-source counts.
\begin{equation}\label{eq:overall_alpha}
\bar{\alpha} = \frac{\displaystyle\sum_{i} \alpha_{i}N_{{\rm OFF},i}}{\displaystyle\sum_{i} N_{{\rm OFF},i}}, \quad i=1,\ldots, m
\end{equation}
Unfortunately, this technique fails to account for any target-specific uncertainties that arise during derivation of the individual $\alpha_{i}$. Even if no systematic \textit{biases} are introduced, the diverse observational parameters that correspond to each individual target dataset will likely affect accuracy to which the corresponding values of $\alpha_{i}$ can be determined \citep{2007A&A...466.1219B}. The resultant non-uniformity renders subsequent propagation of the individual error estimates to the value of $\bar{\alpha}$ somewhat problematic for the data stacking approach. Indeed, it is often unclear how the individual uncertainties should be combined in a statistically consistent manner, particularly when they are characterised by asymmetric intervals or more complicated probability distributions. Assignment of a uniform but conservative systematic error term to all of the $\alpha_{i}$ renders the problem tractable, but inevitably results in an overall loss of sensitivity. 

Moreover, indications exist that the shortcomings of the data stacking technique can be problematic in experimental situations. Indeed, \citet{dickinson_thesis} derived anomalously high combined significance results when applying a data stacking procedure to two independent target ensembles comprising 37 and 44 putative VHE $\gamma$-ray sources for which $N_{{\rm OFF}, i} \sim100-500$ and $\alpha_{i}\sim0.07$. By studying the ON and OFF-source event count distributions, \citet{dickinson_thesis} concluded that a relative target-wise dispersion $\Delta\alpha_{i}/\alpha_{i} \lesssim 0.1$ was required to explain the derived significances in the absence of a true \gr\ signal. 

In subsequent sections, the statistical behaviour of the data stacking approach is investigated, highlighting the limited applicability of this technique in scenarios for which $\alpha$ is uncertain. In response to these inherent shortcomings, a statistically robust method for the combination of IACT observations is outlined, which allows systematic uncertainties to be treated on a target-by-target basis. Using this technique, observations yielding weakly constrained values of $\alpha_{i}$ may be usefully and consistently combined with more reliable datasets, without diluting the scientific utility of the latter. 

\section{Derivation of the Joint Likelihood}\label{sec:joint_like_deriv}

The new method operates by defining target-specific likelihood functions which facilitate appropriate treatment of relevant systematic uncertainties. Stacking is implemented by forming a product of individual target likelihood functions and using this combined likelihood to estimate the shared characteristics of the putative \gr\ signals. It should be emphasised that the applicability of the method is contingent upon an implicit assumption of at least one identical signal characteristic for all targets comprising the sample. Realistically, signal characteristics for which the required assumption of universality is reasonable are typically target-specific functions of the experimental observables. As discussed in the subsequent paragraphs, disparities between putative sub-populations comprising the target sample are inherently problematic in the context of data stacking. In contrast, the suggested approach intrinsically facilitates straightforward customisation of the individual target likelihoods. 

Construction of the single target likelihood function proceeds under the assumption that the sample of \gr-like events that are reconstructed within a nominal sampling region is drawn from a parent population with Poisson distributed arrival times. Accordingly, if the value of $\alpha$ were precisely constrained, then \eqref{eq:N_on_N_off} would imply
\begin{equation}\label{eq:single_mod_dists_1}
N_{\rm ON}\sim\pois(N_{\rm ON},\tilde{N}_{\rm S}+\tilde{\alpha}\tilde{N}_{\rm OFF})\;{\rm and}\; N_{\rm OFF} \sim \pois(N_{\rm OFF},\tilde{N}_{\rm OFF})
\end{equation}
where $\tilde{N}_{\rm ON}$ and $\tilde{N}_{\rm OFF}$ denotes the true mean value of the Poisson distributed variables $N_{\rm ON}$ and $N_{\rm OFF}$ respectively. The new parameter $\tilde{\alpha}$ represents the unknown, true value of the derived normalisation parameter. 
In fact, the assumed validity of \eqref{eq:single_mod_dists_1} is implicit in the derivation of $\bar{\alpha}$ in \eqref{eq:overall_alpha} with likely deviations from this idealised behaviour undermining the reliability of the traditional data stacking framework. Conversely, the combined likelihood approach treats target-specific systematic uncertainties by modelling $\alpha$ as a random variable with arbitrary distribution $G(\alpha|\tilde{\alpha})$. As indicated in $\S$\ref{subsec:obs_technique}, the specific functional definitions of $G(\alpha|\tilde{\alpha})$ are dependent upon observational parameters which are likely to vary on a target-by-target basis. 

The probability density function which models an abstract single-target dataset $D=\{N_{\rm ON}, N_{\rm OFF}, \alpha\}$  is simply the product of the distributions of its component data.
\begin{align}
f_{1}(N_{\rm ON}, N_{\rm OFF}, \alpha |\tilde{N}_{\rm S},\tilde{N}_{\rm OFF},\tilde{\alpha})&= \pois(N_{\rm ON}, \tilde{N}_{\rm S}+\tilde{\alpha} \tilde{N}_{\rm OFF})\notag\\ 
&\qquad\cdot\pois(N_{\rm OFF}, \tilde{N}_{\rm OFF})\label{eq:single_mod_pdf_3}\\
&\qquad\;\;\,\cdot G(\alpha|\tilde{\alpha})\notag
\end{align}
Accordingly, substitution for $D$ using a concrete observational dataset $d=\{n_{\rm ON}, n_{\rm OFF}, \alpha_{\rm obs}\}$ enables calculation of the target-specific likelihood describing the true \gr\ signal.
\begin{equation}
L_{1}(\tilde{N}_{S},\tilde{N}_{\rm OFF},\tilde{\alpha} | n_{\rm ON}, n_{\rm OFF}, \alpha_{\rm obs}) =f_{1}(n_{\rm ON}, n_{\rm OFF}, \alpha_{\rm obs} |\tilde{N}_{\rm S},\tilde{N}_{\rm OFF},\tilde{\alpha})\label{eq:single_mod_pdf_4}
\end{equation}
Modelling systematically uncertain parameters as random variables with known probability density functions is a common statistical approach, with applications extending beyond the limited domain of stacking analyses \citep[e.g.][]{Heinrich:2007zza}. Indeed, equations \eqref{eq:single_mod_pdf_3} and \eqref{eq:single_mod_pdf_4} are equally valid in single target IACT analyses that must accommodate non-negligible systematic uncertainties on the estimated value of $\alpha$.

The ultimate goal of this stacking procedure is to obtain improved constraints on the global \gr\ signal characteristics. In this context, $\tilde{N}_{\rm OFF}$ and $\tilde{\alpha}$ are not of primary interest and are therefore categorised as nuisance parameters throughout the subsequent analysis . 

Finally, the joint likelihood describing the global signal characteristics of the entire ensemble of $m$ targets is simply the product of the individual target likelihoods.
\begin{equation}\label{eq:joint_mod_lhood}
L_{{\rm J},m}(\tilde{N}_{\rm S})=\prod_{i=1}^{m}{L_{1,i}(\tilde{N}_{\rm S},\tilde{N}_{{\rm OFF},i},\tilde{\alpha}_{i}|n_{{\rm ON},i},n_{{\rm OFF},i},\alpha_{{\rm obs},i})}
\end{equation}
Accordingly, the set of values which parameterise $L_{\rm J}$ is the union of the parameters of the component single-target likelihood functions and therefore incorporates $2m$ uncertain nuisance parameters.

\section{Inference of the Signal Characteristics}\label{sec:inf_sig_char}

\subsection{Detection Significance}

The initial objective of a stacking analysis is typically the conclusive detection of a combined \gr\ excess, with detailed investigation of specific signal characteristics assuming subsidiary importance. In such situations, the probability of detection is typically evaluated in terms of the the significance, $S$ with which experimental data exclude the null hypothesis of zero signal.  

For an abstract likelihood function $L(\boldsymbol{\pi}, \boldsymbol{\theta}|D)$ relating the experimental data $D=\{D_{1}\ldots D_{n}\}$ to multiple nuisance parameters, $\boldsymbol{\theta} = \{\theta_{1},\ldots,\theta_{j}\}$ and parameters of interest $\boldsymbol{\pi} = \{\pi_{1},\ldots,\pi_{l}\}$, a popular formula for calculation of detection significance \citep{1983ApJ...272..317L} employs the likelihood ratio defined by \citep{Neyman01011933}
\begin{equation}\label{eq:li_ma_ratio}
\lambda_{\rm LM}=\frac{L(\boldsymbol{\pi}_{0}, \hat{\boldsymbol{\theta}}_{0}|D)}{L(\hat{\boldsymbol{\pi}}, \hat{\boldsymbol{\theta}}|D)}
\end{equation}
where a caret denotes the maximum likelihood estimate of the accented symbol, $\boldsymbol{\pi}_{0}$ denotes the value of $\boldsymbol{\pi}$ which corresponds to the null hypothesis, and $\hat{\boldsymbol{\theta}}_{0}$ represents the conditional maximum likelihood estimate for $\boldsymbol{\theta}$ given that $\boldsymbol{\pi}=\boldsymbol{\pi}_{0}$. The significance is subsequently defined in terms of $\lambda_{\rm LM}$
\begin{equation}\label{eq:li_ma_sig}
S=\sqrt{-2\ln\lambda_{\rm LM}}
\end{equation}
which reduces to
\begin{equation}\label{eq:LiMa17}
S=\sqrt{2}\left\{N_{\rm ON}\ln\left[\left(\frac{1+\alpha}{\alpha}\right)\frac{N_{\rm ON}}{N_{\rm ON} + N_{\rm OFF}} \right]^{N_{\rm ON}} + N_{\rm OFF}\ln\left[(1+\alpha)\frac{N_{\rm OFF}}{N_{\rm ON} + N_{\rm OFF}} \right]^{N_{\rm ON}}\right\}^{\frac{1}{2}}
\end{equation}
as reported by \cite[][Equation 17]{1983ApJ...272..317L} and is nominally $\chi_{1}$ distributed for a typical, single-target dataset with negligible $\alpha$ uncertainty i.e. $D=\{N_{\rm ON}, N_{\rm OFF}, \alpha\}$. Proponents of data stacking typically emulate the single-target approach by using Equation 17 in conjunction with a pseudo-dataset $D^{\prime}=\{\sum_{i}{N_{{\rm ON},i}}, \sum_{i}{N_{{\rm OFF},i}}, \bar{\alpha}\}$. Consequently, non-uniform observational dependencies the individual $\alpha_{i}$ become entangled, which inevitably complicates statistically robust treatment of their effects. Conversely, direct substitution of $L_{J}$ in the calculation of $\lambda_{LM}$ in the joint-likelihood approach ensures that the resultant expression for $S$ is constructed using all available information regarding target-specific systematic uncertainties.

Recently, \citet{2011arXiv1112.0786K} proposed an extension to the method suggested by \citet{1983ApJ...272..317L} which uses well-calibrated estimates of the instrumental point-spread function to achieve enhanced sensitivity relative to the more traditional analysis. Their technique ameliorates the effect of systematic variations in the telescope acceptance characteristics by segregating the observational data on the basis of prevailing instrumental and environmental parameters. 

\subsection{Confidence Intervals}

Results that are derived from a joint likelihood stacking analysis comprise estimates of the various experimental parameters of interest, with any uncertainty quantified using associated confidence intervals. The frequentist definition of a properly constructed $100(1-\epsilon)\%$ confidence interval states that $100(1-\epsilon)\%$ of such intervals, generated from a large number of independent experimental measurements of a quantity $X$, will contain (or \emph{cover}) the true value $\tilde{X}$. Confidence intervals which fulfil this criterion are described as having correct \emph{coverage}. Construction of confidence intervals typically uses experimental data to define a test statistic which is subsequently used to identify regions of the relevant parameter space that fulfil the required statistical criteria. The \emph{profile likelihood} is an appropriate choice of test statistic in situations that require the treatment of multiple nuisance parameters. The profile likelihood is defined as the ratio
\begin{equation}\label{eq:prof_like}
\lambda_{\rm PL}(\boldsymbol{\pi}| D) = \frac{L(\boldsymbol{\pi},\hat{\boldsymbol{\theta}}_{\pi}|D)}{L(\hat{\boldsymbol{\pi}},\hat{\boldsymbol{\theta}}|D)}
\end{equation}
where previously encountered symbols retain their earlier definitions and $\hat{\boldsymbol{\theta}}_{\pi}$ represents the conditional maximum likelihood estimate for $\boldsymbol{\theta}$ assuming a specific value of $\boldsymbol{\pi}$. Conveniently, the distribution of $-2\log\lambda_{\rm PL}$ converges to that of a $\chi^{2}$ random variate with $l$ degrees of freedom for large experimental data sets \cite[e.g.][]{casella2002statistical}. Accordingly, straightforward derivation of confidence intervals is facilitated by comparison of the derived test statistic with appropriate percentiles of the relevant $\chi^{2}$ distribution. Moreover, the profile likelihood has the desirable property of being independent of $\boldsymbol{\theta}$, facilitating the derivation of robust confidence intervals, even in the presence of uncertain nuisance parameters. 

\section{Monte Carlo Studies}\label{sec:monte_carlo}

\begin{figure}[t]
\begin{center}
\includegraphics[width=0.6\textwidth]{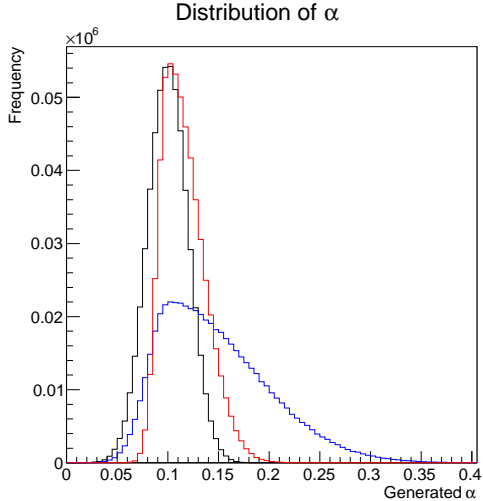}
\end{center}
\caption{Generated Monte Carlo distributions of the ON/OFF normalisation $\alpha$ corresponding to Model A (\emph{black}), Model B (\emph{red}), and Model C (\emph{blue}).}\label{fig:alphadists}
\end{figure}

Verification of the performance and reliability joint likelihood stacking procedure employed a computerised toy Monte Carlo approach. The RooFit\footnote{http://roofit.sourceforge.net} framework \citep{2006sppp.conf..186V} was used to generate multiple simulated datasets consistent with the assumed distributions of $N_{\rm ON}$, $N_{\rm OFF}$ and $\alpha$ for specific values of the true signal parameters $\tilde{N}_{\rm S}\in[0, 10]$, $\tilde{N}_{\rm OFF} = 100$ and $\tilde{\alpha} = 0.1$. To better understand the effect of the distribution of $\alpha$ on the outcome of each stacking procedure, data were generated using three alternative parameterisations for $G(\alpha|\tilde{\alpha}_{i})$. A basic test case, simulating unbiassed systematic uncertainties, employed a Gaussian parameterisation $G(\alpha|\tilde{\alpha})={\rm Gaus}(\tilde{\alpha}, \sigma_{\alpha})$ with $\sigma_{\alpha} = 0.02$ (hereafter referred to as \emph{Model A}). Two less idealised examples, adopting bifurcated Gaussian functions
\begin{equation}\label{eq:bifugauss}
G(\alpha|\tilde{\alpha})=\left\{\begin{array}{rl}{\rm Gaus}(\tilde{\alpha}, \sigma_{\alpha,{\rm L}}) & {\rm if}~\alpha < \tilde{\alpha} \\
{\rm Gaus}(\tilde{\alpha}, \sigma_{\alpha,{\rm R}}) & {\rm if}~\alpha > \tilde{\alpha} \\
\end{array}\right.
\end{equation}
with $(\sigma_{\alpha,{\rm L}}, \sigma_{\alpha,{\rm R}})=(0.01, 0.03)$ and $(\sigma_{\alpha,{\rm L}}, \sigma_{\alpha,{\rm R}})=(0.02, 0.08)$ (hereafter referred to as \emph{Model B} and \emph{Model C}, respectively), allowed the effect of increasing asymmetry in the modelled $\alpha$ distribution to be investigated. Figure \ref{fig:alphadists} shows representative distributions of $\alpha$ corresponding to each of the simulated scenarios. Models A and B mimic somewhat pessimistic observational situations in which the choice of OFF-source sampling regions is restricted, limiting event statistics and biasing the derivation of $\alpha$. Accordingly, the corresponding results are particularly relevant to galactic source populations, for which source confusion and diffuse background contamination within target fields-of-view can be particularly problematic. Model C represents an extreme case which is unlikely to occur in real experimental contexts, but is nonetheless included to demonstrate the efficacy of the joint likelihood technique.  

Using RooFit, ensembles of the generated single-target datasets were then employed in conjunction with the relevant compound probability density functions to obtain multiple realisations of the joint likelihood $L_{{\rm J},m}$ corresponding to a discrete range of target multiplicities $m\in[1,10]$. For each ensemble dataset $D_{m}$, the statistical significance ($S_{\rm J}$) of the combined \gr\ signal was evaluated using the likelihood ratio prescription of \citep{1983ApJ...272..317L}
\begin{equation}
S_{\rm J} = -2\log\left\{\frac{L_{{\rm J},m}(N_{\rm S} = 0, \hat{\boldsymbol{\theta}}_{0}|D_{m})}{L_{{\rm J},m}(\hat{N}_{\rm S},\hat{\boldsymbol{\theta}}|D_{m})}\right\},~\hat{\theta}_{i}=\{\hat{N}_{{\rm OFF},i},\hat{\alpha}_{i}\},~i=1\ldots m.
\end{equation}

For comparison, a traditional data stacking analysis was also applied to the generated data. The components of the ensemble datasets used to construct each realisation of $L_{{\rm J},m}$ were combined to obtain corresponding stacked datasets, $D^{\prime}_{m}=\{\sum_{i}{N_{{\rm ON},i}}, \sum_{i}{N_{{\rm OFF},i}}, \bar{\alpha}\}$. Subsequent statistical inference employed a likelihood function $L_{\rm DS}$, which was formed by substitution of the stacked data into the single target probability density function $f_{1}$ with $G$ defined as a delta function centred on the stacked $\bar{\alpha}$ estimate, $G(\alpha|\tilde{\alpha}, \bar{\alpha})=\delta(\alpha - \bar{\alpha})$. Following the convention outlined in $\S$\ref{sec:inf_sig_char}, Equation \eqref{eq:LiMa17} was evaluated using the components of $D^{\prime}_{m}$ do derive the statistical significance $S_{\rm DS}$ of the combined \gr\ signal corresponding to the data stacking approach.

For both stacking methodologies, construction of the associated profile likelihoods was accomplished using the intrinsic capabilities of RooFit, adopting $\tilde{N}_{\rm S}$ as the sole parameter of interest. Confidence intervals were then defined as the ranges of $\tilde{N}_{\rm S}$ outside which $-2\log\lambda_{\rm PL}$ exceeded an appropriate percentile of the nominal $\chi^{2}_{1}$ distribution. 

\section{Comparison of the Alternative Stacking Techniques}\label{sec:results}

For each distinct combination of the true signal parameters, modelled distribution of $\alpha$, and target multiplicity, the Monte Carlo approach outlined in $\S$\ref{sec:monte_carlo} was applied to derive 5000 independent estimates of $N_{S}$ and $S$ corresponding to each stacking technique. In combination with appropriate statistical metrics, the resultant data enable a comparative evaluation of the joint likelihood and traditional data stacking approaches.

\subsection{Detection Significance}

An experimentally relevant diagnostic for both stacking analyses is the probability with which data corresponding to a particular value of $\tilde{N}_{\rm S}$ will yield a significance $S$, in excess of a specified threshold $S_{\rm t}$. Assuming data that are distributed in accordance with the null hypothesis, this probability is called the \emph{significance level}. Although the specific choice of $S_{\rm t}$ is arbitrary, it is typically chosen in order to obtain a desired significance level and ensure a controlled rate of false-positive detections. Objective comparison of the alternative stacking techniques is facilitated by a threshold which corresponds to a nominal percentile of the appropriate null distribution of $S$. A related statistic is the \emph{power}, which describes the probability that derived values of $S>S_{\rm t}$ will correctly identify a genuine signal, and provides useful insights regarding the sensitivity of the corresponding analytical method. 

\begin{figure}[h]
\begin{center}
\includegraphics[width=0.45\textwidth]{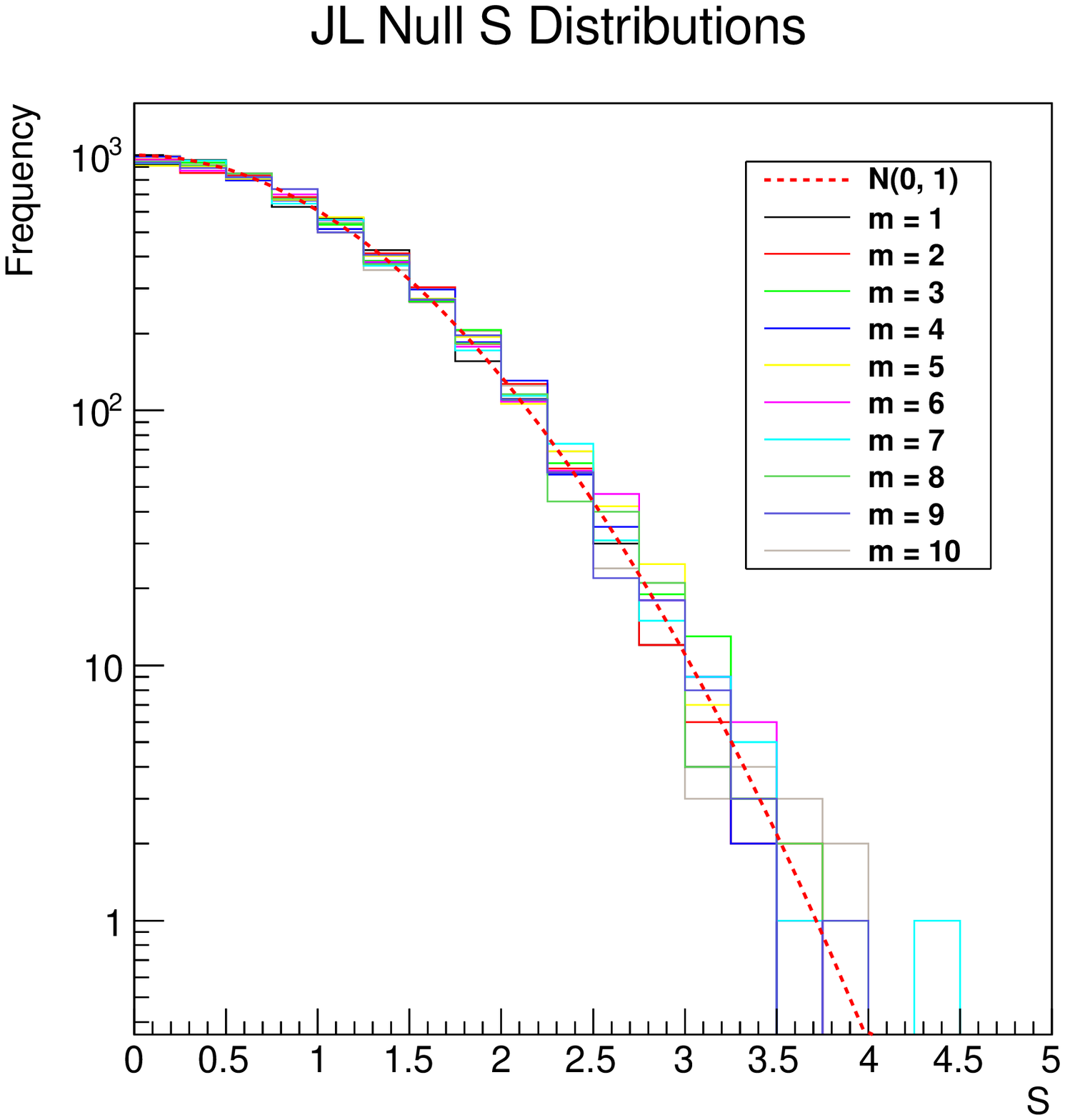}
\includegraphics[width=0.45\textwidth]{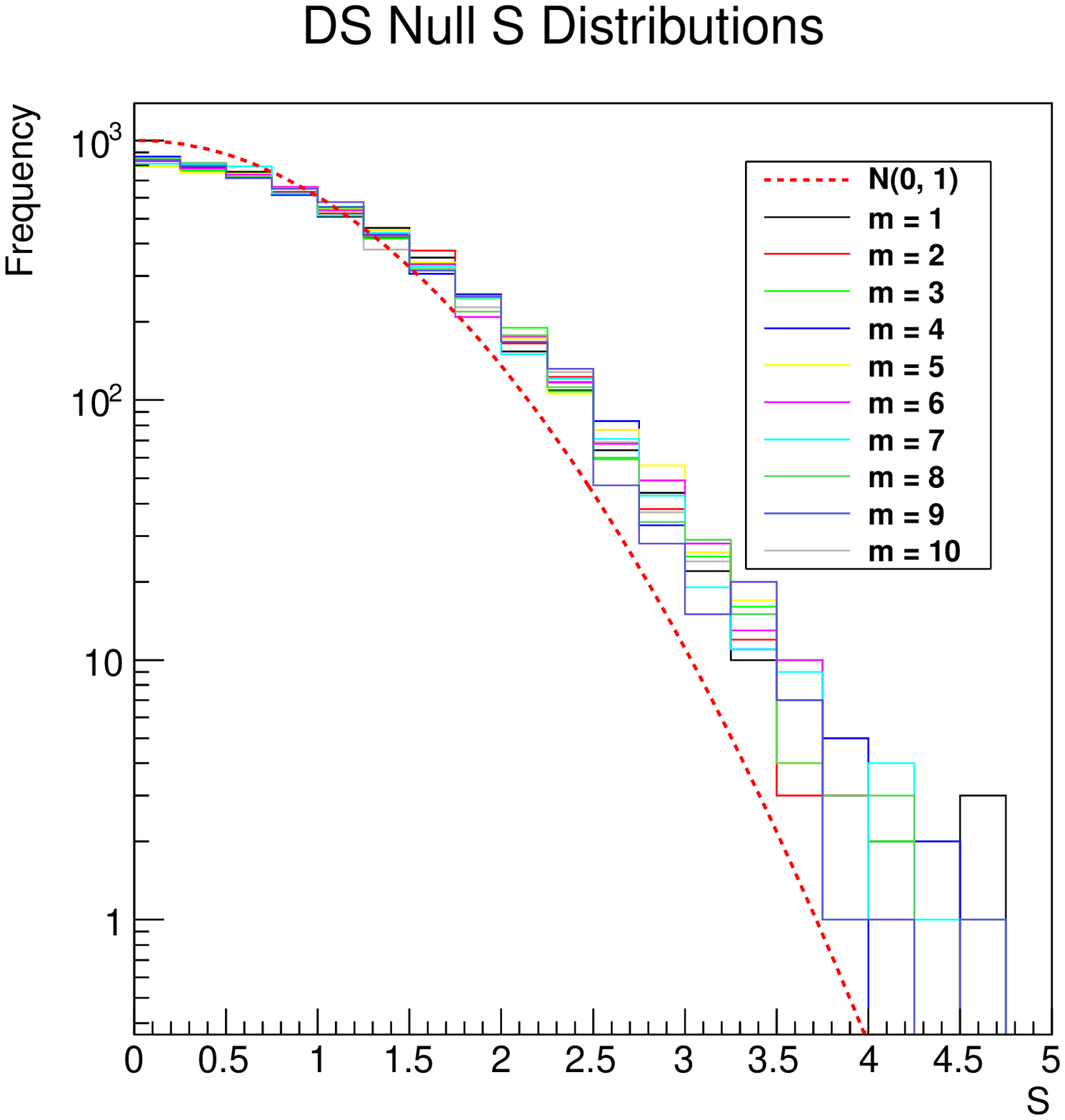}\\
\includegraphics[width=0.45\textwidth]{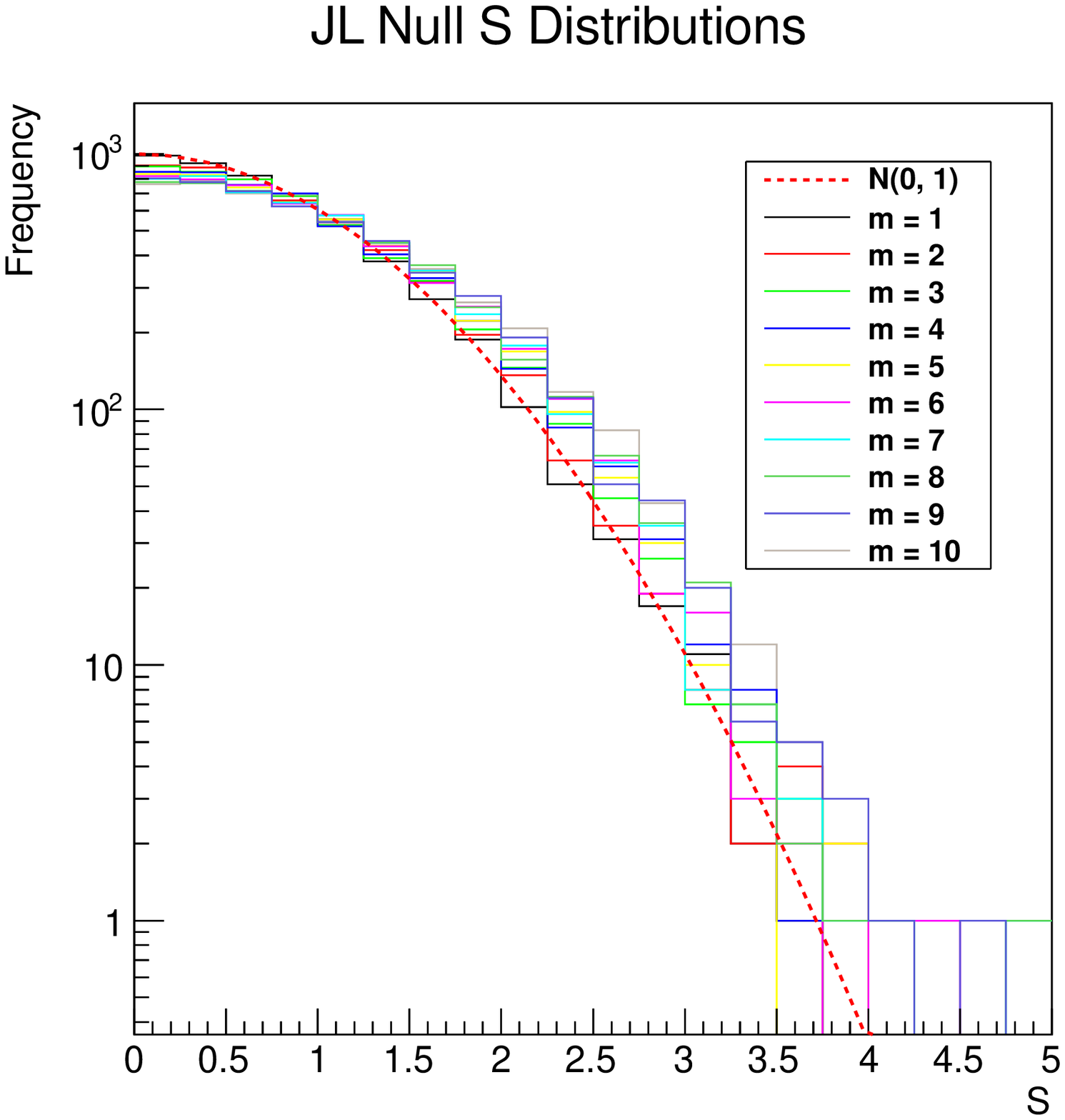}
\includegraphics[width=0.45\textwidth]{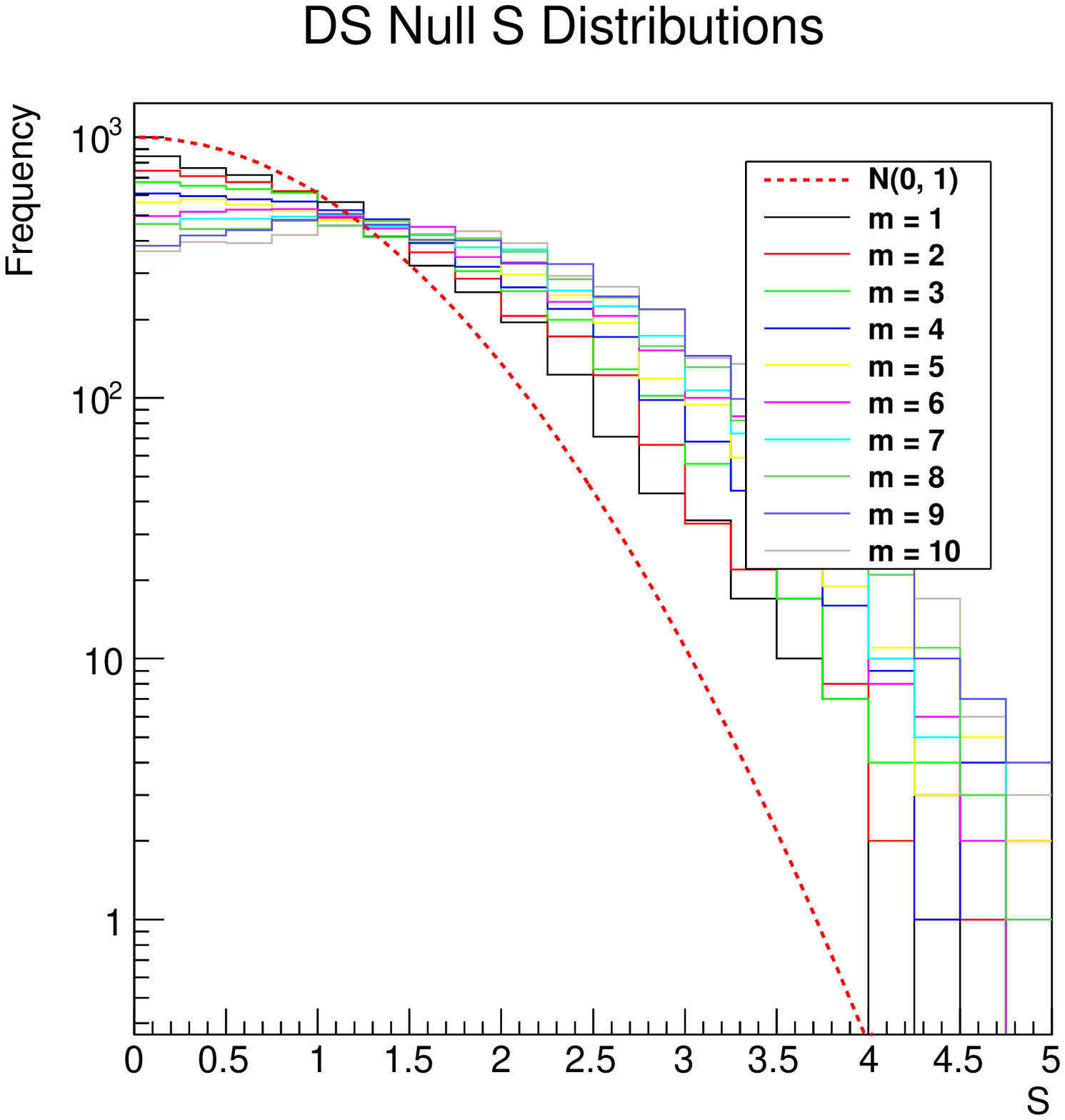}\\
\includegraphics[width=0.45\textwidth]{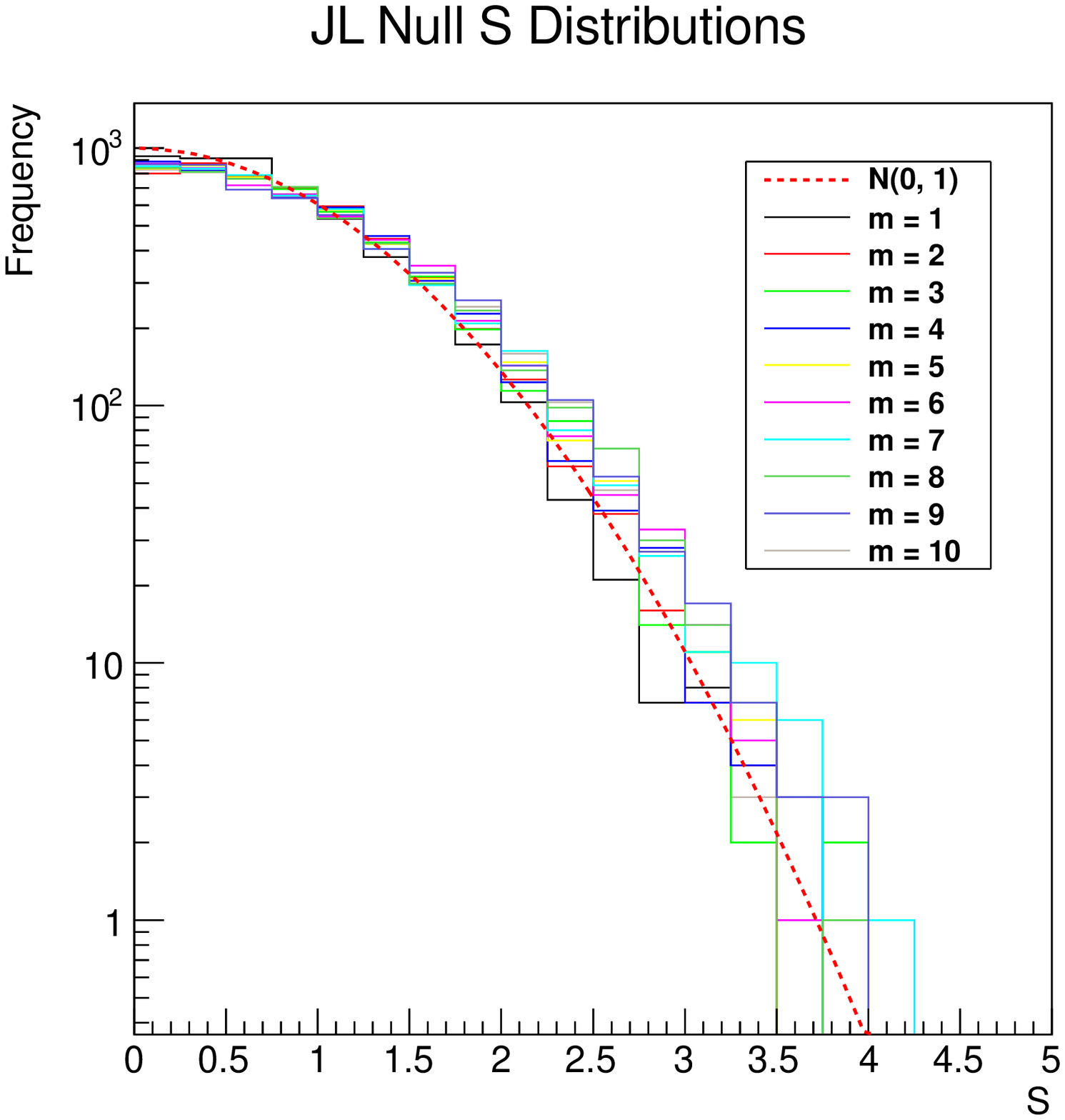}
\includegraphics[width=0.45\textwidth]{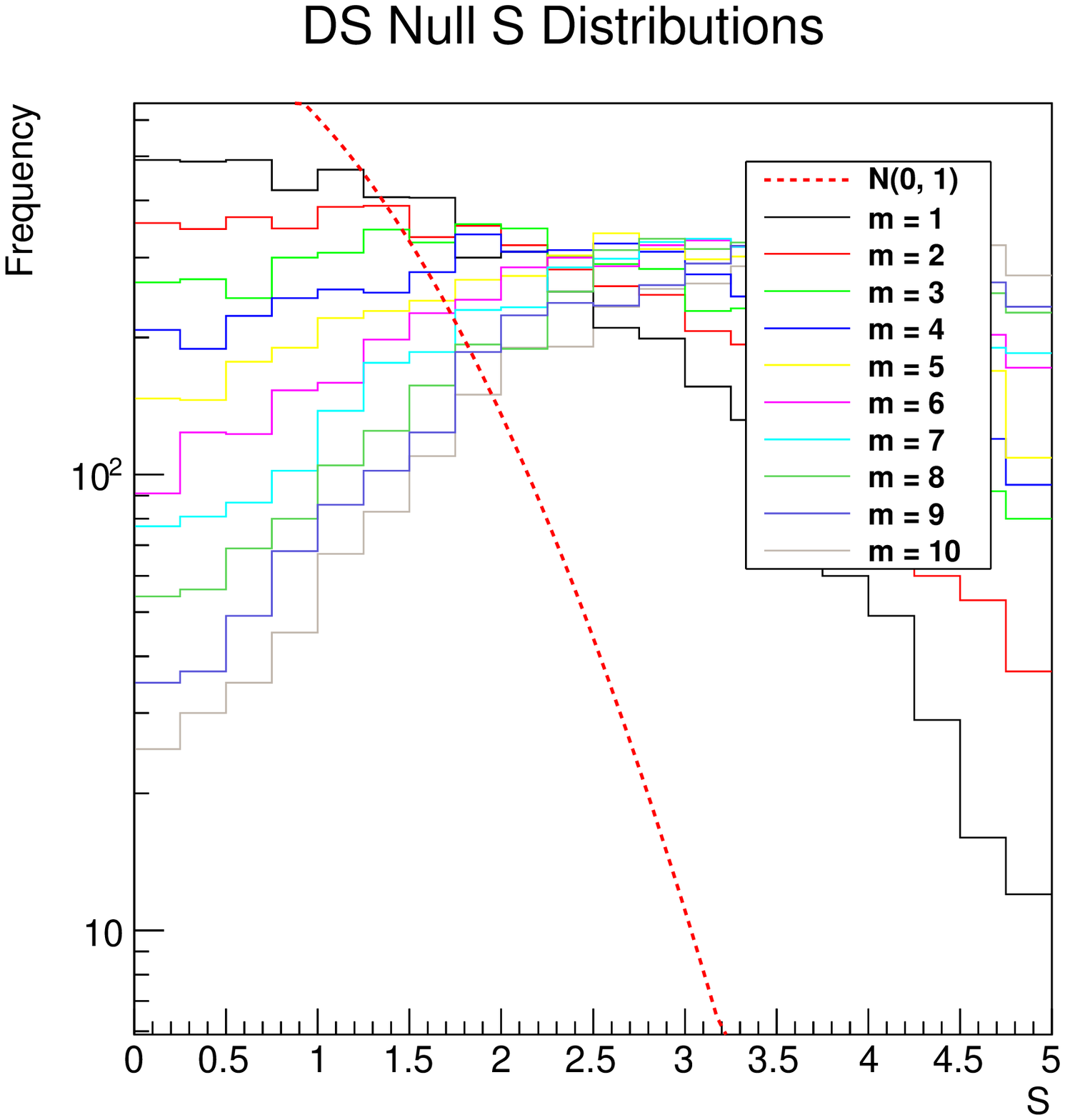}
\end{center}
\caption{Monte Carlo distributions for $S$ corresponding to the Joint Likelihood and Data Stacking approaches. Each histogram is representative of 5000 independent Monte Carlo datasets, which were generated assuming $\tilde{N}_{\rm S}=0$. Results are shown which correspond Model A (\emph{top row}), Model B (\emph{middle row}) and Model C (\emph{bottom row}) for target multiplicities $m\in[1,10]$. All Monte Carlo datasets were generated assuming $\tilde{N}_{\rm OFF} = 100$ and $\tilde{\alpha} = 0.1$.}\label{fig:null_sigma_dists}
\end{figure}

The relevant Monte Carlo distributions for $S$, corresponding to $\tilde{N}_{\rm S}=0$, are plotted for each target multiplicity in Figure \ref{fig:null_sigma_dists}. 
Assuming that the true value of $\tilde{\alpha}$ is 0.1, systematic uncertainties which affect the estimation of $\alpha_{i}$ according to the distributions plotted in Figure \ref{fig:alphadists} will generate misleading results in a data stacking analysis. 
In many experimental situations, a small number of erroneous results is tolerable, as long as the expected frequency with which they occur can be reliably predicted. This predictability criterion is realised if the derived values consistently conform to a known statistical distribution. Irrespective of $G(\alpha|\tilde{\alpha})$, and for all $m\in[0, 10]$, the significance estimates derived using the joint likelihood technique appear to retain the $\chi_{1}$ distribution which arises in case of zero $\alpha$ uncertainty. In contrast, results that correspond to traditional data stacking exhibit substantial deviations from this nominal distribution. Moreover, adoption of asymmetric $\alpha$ distributions introduces a monotonic dependence on the target multiplicity, yielding systematically larger significance estimates as $m$ increases.

Consideration of the procedure used during data stacking to synthesise a combined $\alpha$ estimate suggests an intuitive explanation of the observed behaviour. Models B and C generate ensemble data sets in which the subset of overestimated $\alpha$ values is typically predominant and has an absolute \textit{mean} offset from $\tilde{\alpha}$ which exceeds that of the remaining, underestimated $\alpha_{i}$. Accordingly, the frequency with which Equation \eqref{eq:overall_alpha} yields a substantial overestimate for $\bar{\alpha}$ should increase with the ensemble size, which is consistent with the observed behaviour. 

Conversely, symmetrically distributed $\alpha$ uncertainties should not inherently bias the data stacking process. Instead, the target-specific weighting of each $\alpha_{i}$ specified in Equation \eqref{eq:overall_alpha} prevents perfect cancellation of the systematic offsets and inhibits convergence to the nominal null distribution for $S$ as $m\rightarrow\infty$. As outlined in $\S$\ref{sec:monte_carlo} all Monte Carlo data sets used for this study assume a common $\tilde{N}_{\rm OFF} = 100$, with the range of inter-target variability in $N_{{\rm OFF},i}$ restricted to the level of Poisson fluctuations. In experimental contexts, the true number of OFF events is likely to be different for each target and the weights accorded to the corresponding $\alpha_{i}$ would be more diverse. The adverse impact of symmetric systematic uncertainties would then depend critically on the joint, target-wise distribution of $\alpha$-offset and $N_{{\rm OFF},i}$, with specific ensembles potentially realising particularly benign or pathological scenarios.

As an alternative to the absolute measure of significance defined by Equation \eqref{eq:LiMa17}, experimental practitioners often choose to multiply $S$ by the sign of the corresponding \gr\ excess. Adopting this convention, systematic \textit{over}estimates of $\bar{\alpha}$ during data stacking would tend to over-generate negative significances when $\tilde{N}_{\rm S}=0$. In many situations the notion of a negative signal is not physically meaningful and such results would probably be disregarded. In general, the ability to eliminate spurious results on the basis of their physical validity may ameliorate the impact of particular systematic effects, and reduce the rate of false-positive detections. Nonetheless, unfeasibly negative significances should raise concerns regarding the trustworthiness of corresponding upper limits. 

\begin{figure}
\begin{center}
\includegraphics[width=0.45\textwidth]{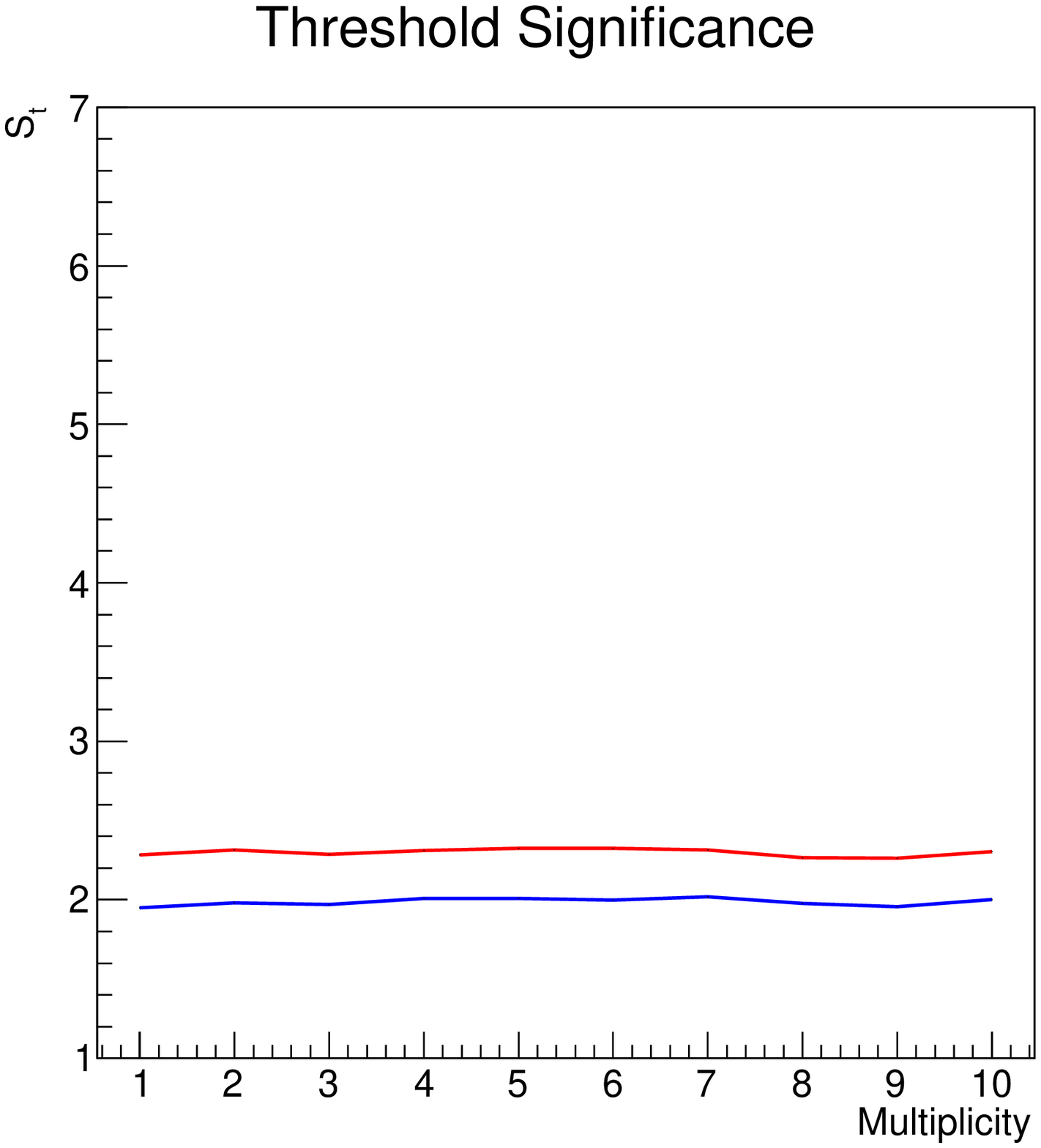}
\includegraphics[width=0.45\textwidth]{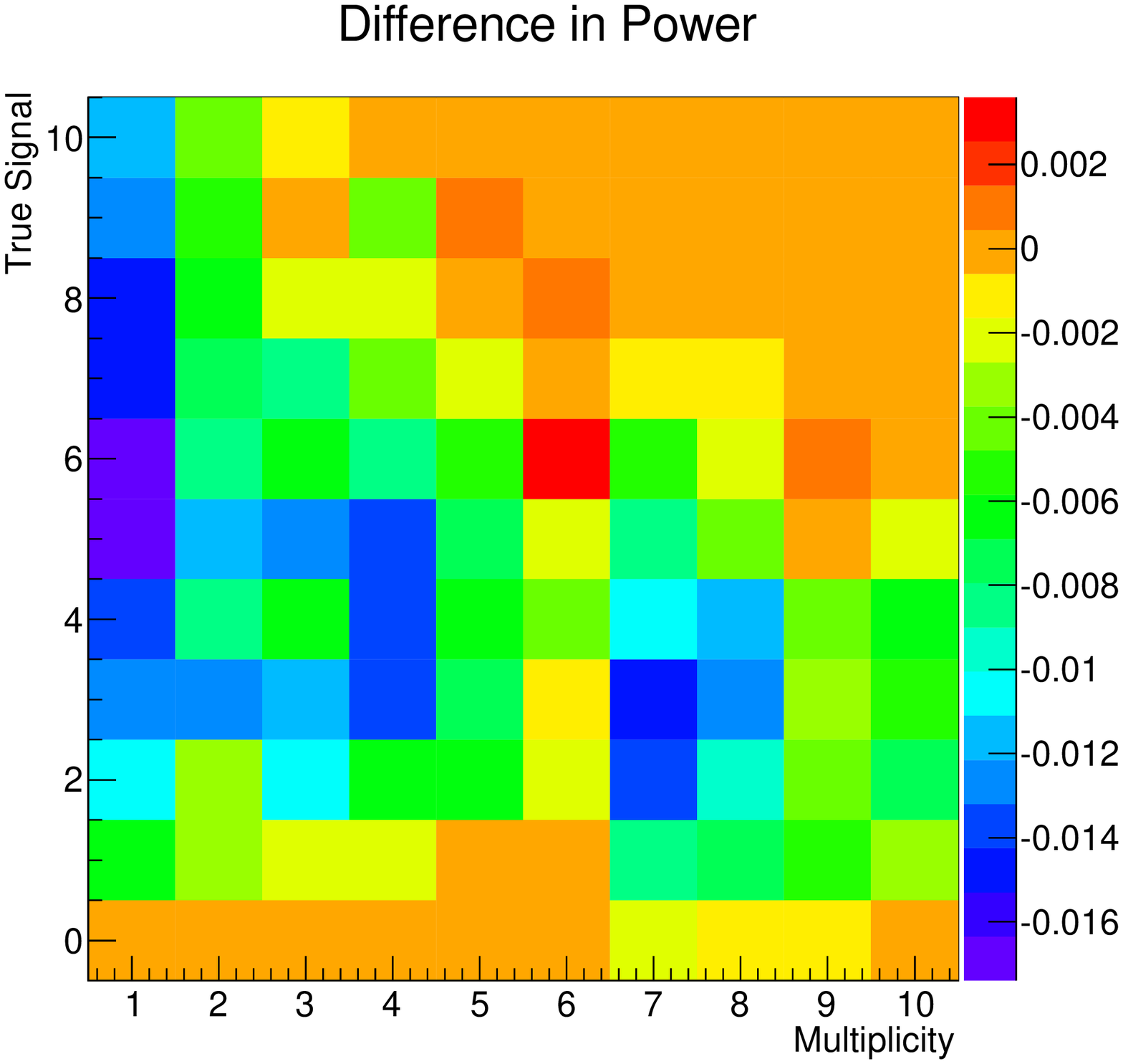}\\
\includegraphics[width=0.45\textwidth]{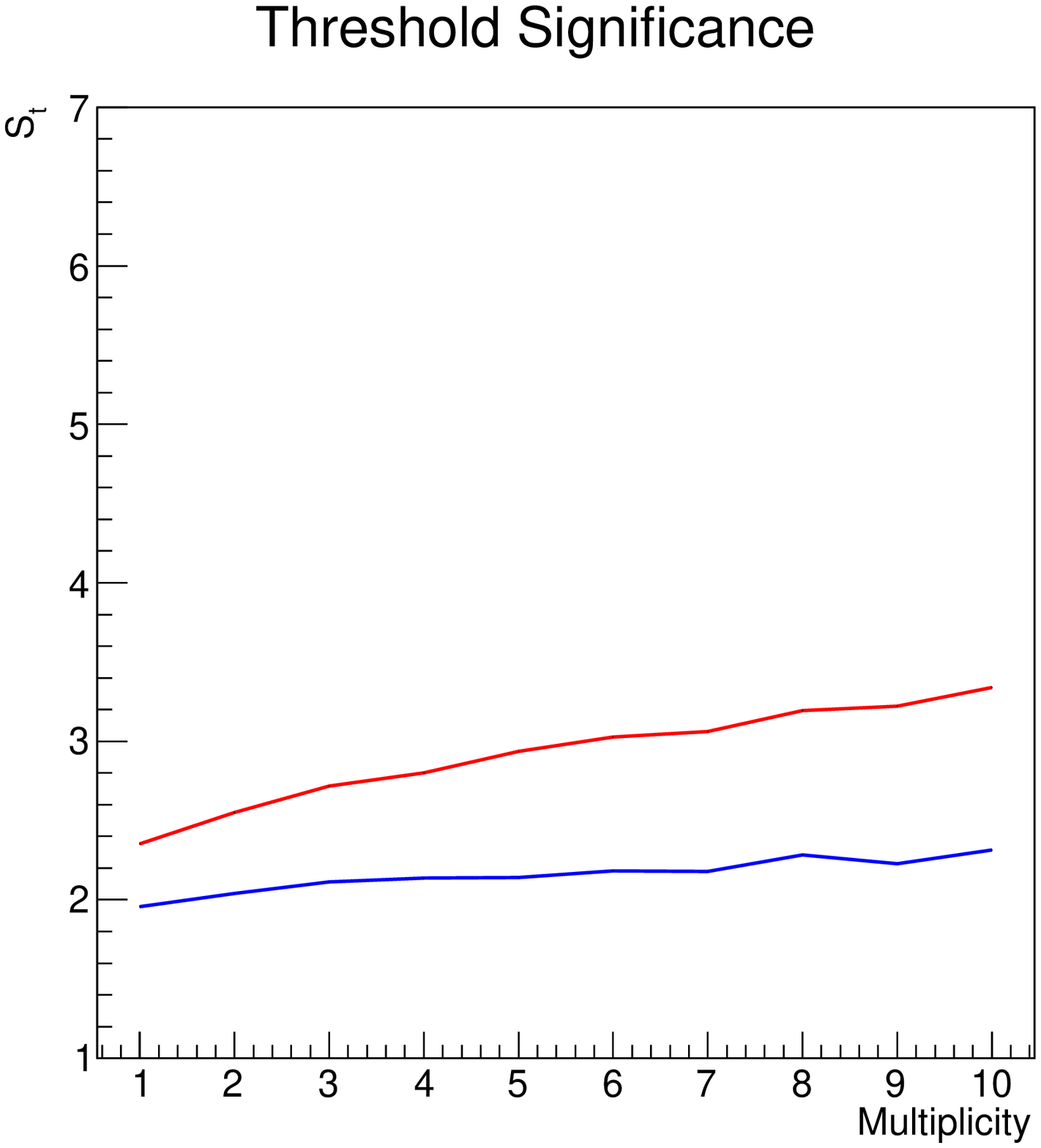}
\includegraphics[width=0.45\textwidth]{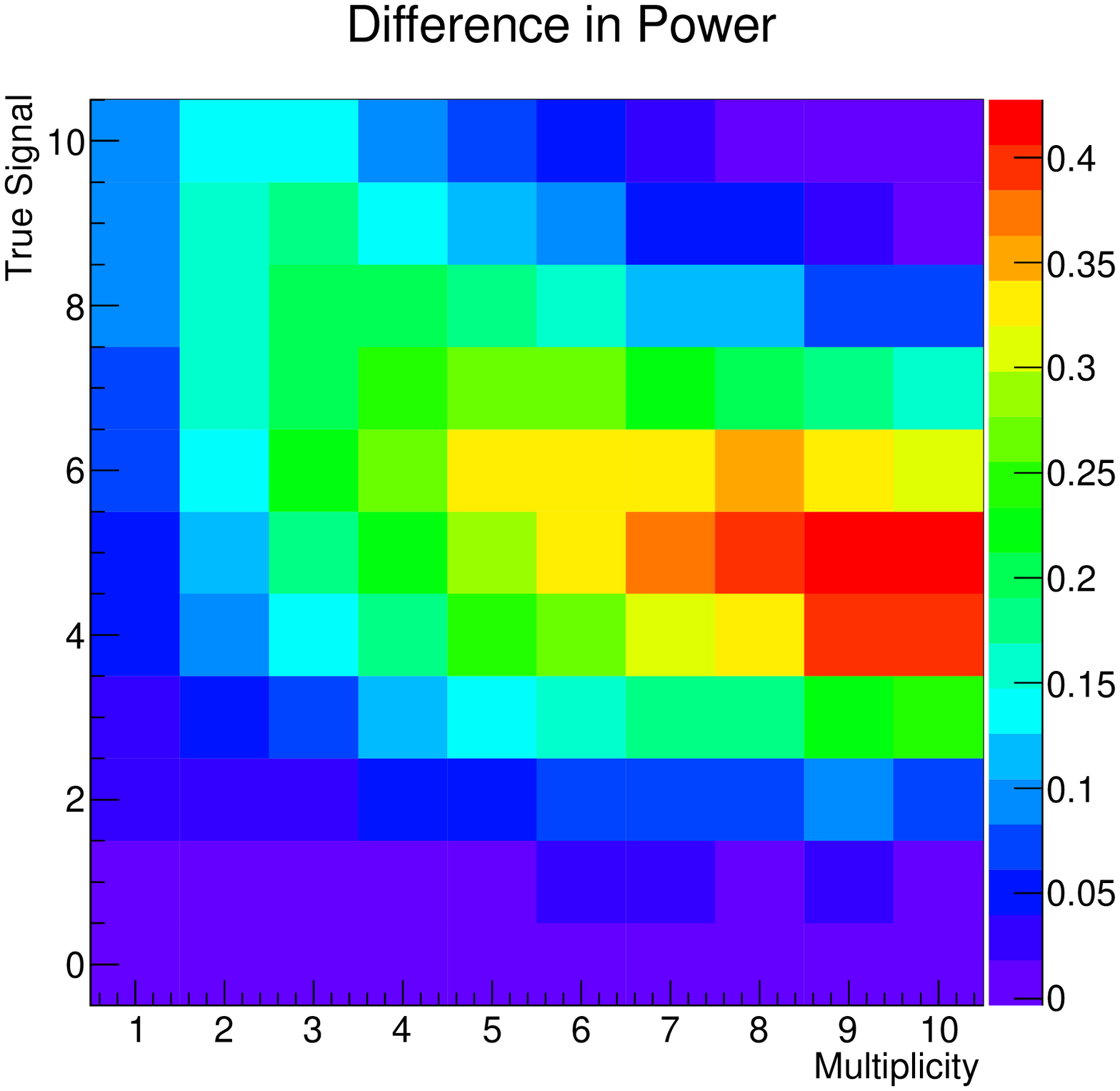}\\
\includegraphics[width=0.45\textwidth]{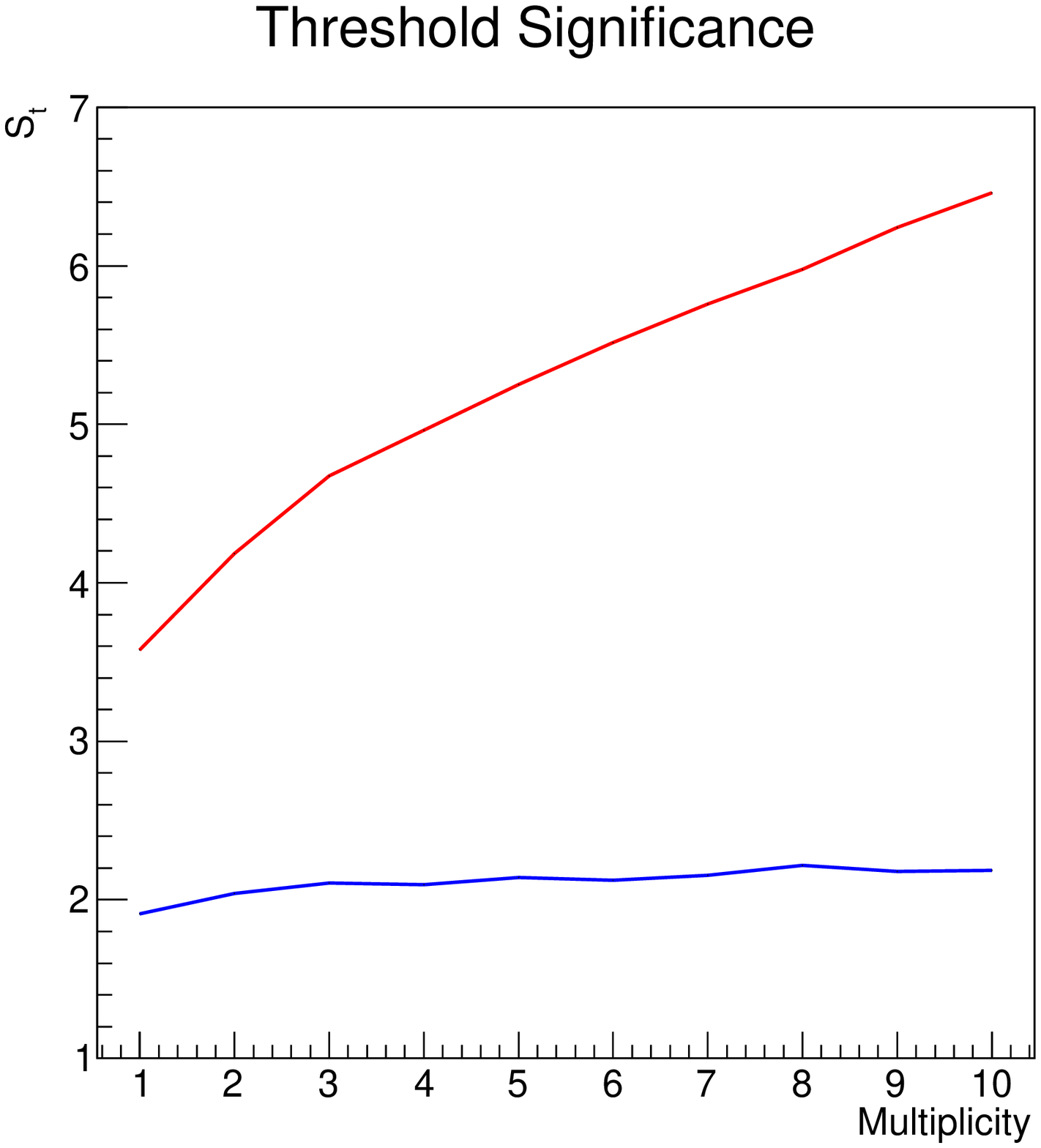}
\includegraphics[width=0.45\textwidth]{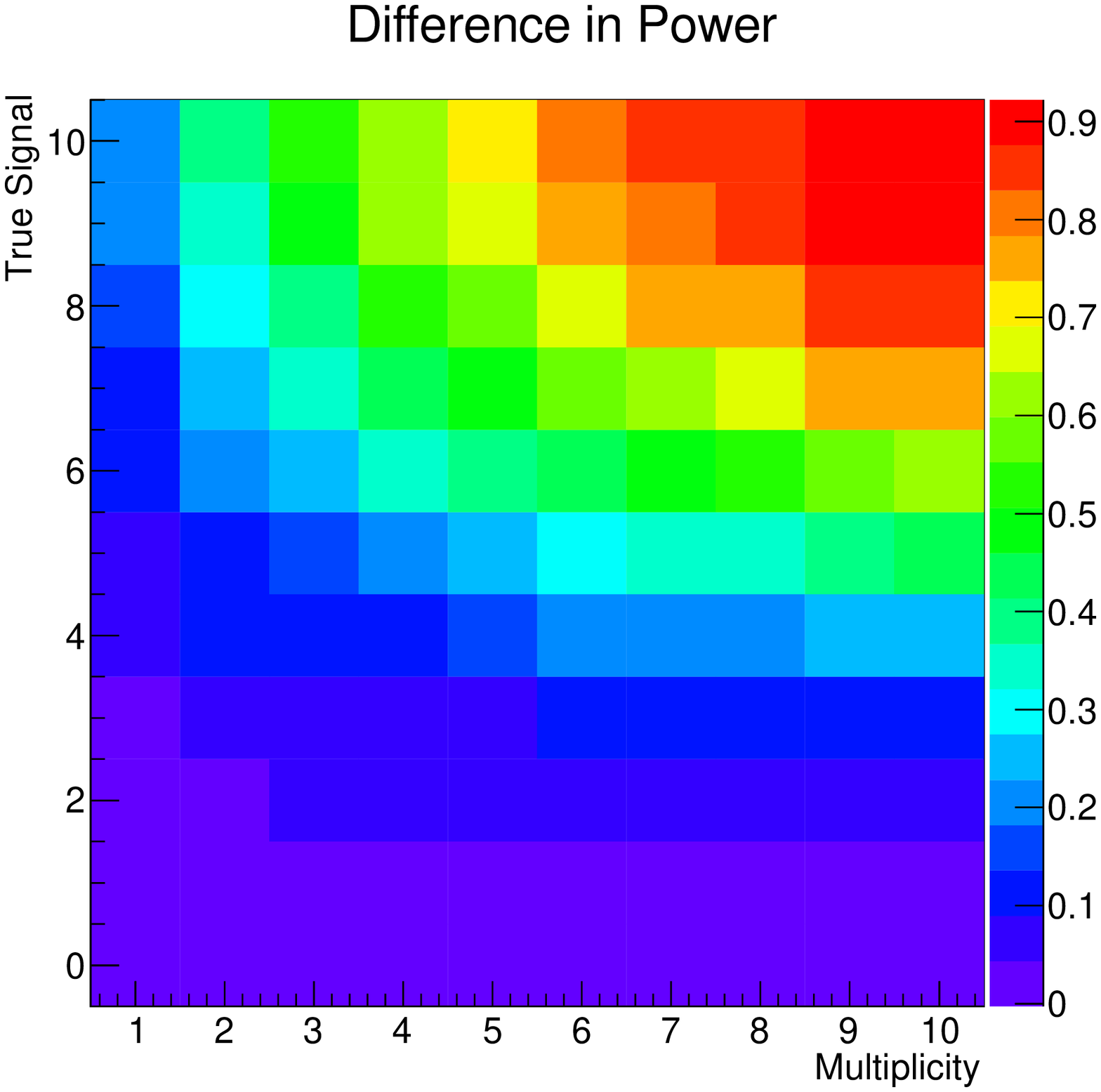}
\end{center}
\caption{\emph{Left-hand column:} Calibrated significance threshold values corresponding to the 95th percentile of the Monte Carlo null distributions of $S$ for the Joint Likelihood (\emph{blue}) and Data Stacking (\emph{red})  approaches. \emph{Right-hand column:} Differences between the derived powers corresponding to the Joint Likelihood and Data Stacking approaches. The plotted results correspond to Model A (\emph{top row}), Model B (\emph{middle row}) and Model C (\emph{bottom row}).  Each bin is derived using analyses of 5000 independent Monte Carlo datasets which assume to true signals $\tilde{N}_{\rm S}\in[0, 10]$ and target multiplicities $m\in[1,10]$. All Monte Carlo datasets were generated assuming $\tilde{N}_{\rm OFF} = 100$ and $\tilde{\alpha} = 0.1$.}\label{fig:power_diffs}
\end{figure}

\begin{figure}
\begin{center}
\includegraphics[width=0.45\textwidth]{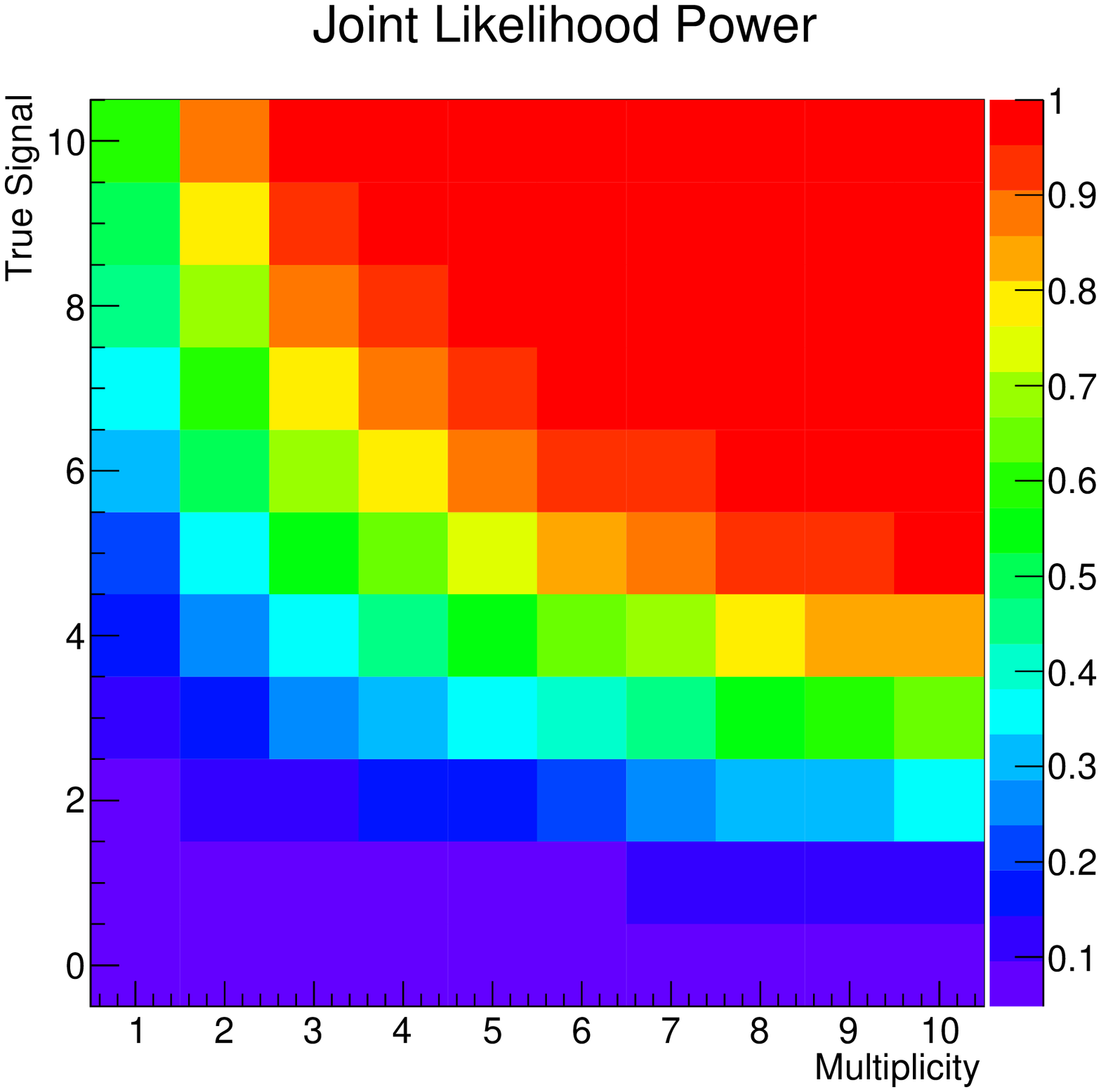}
\includegraphics[width=0.45\textwidth]{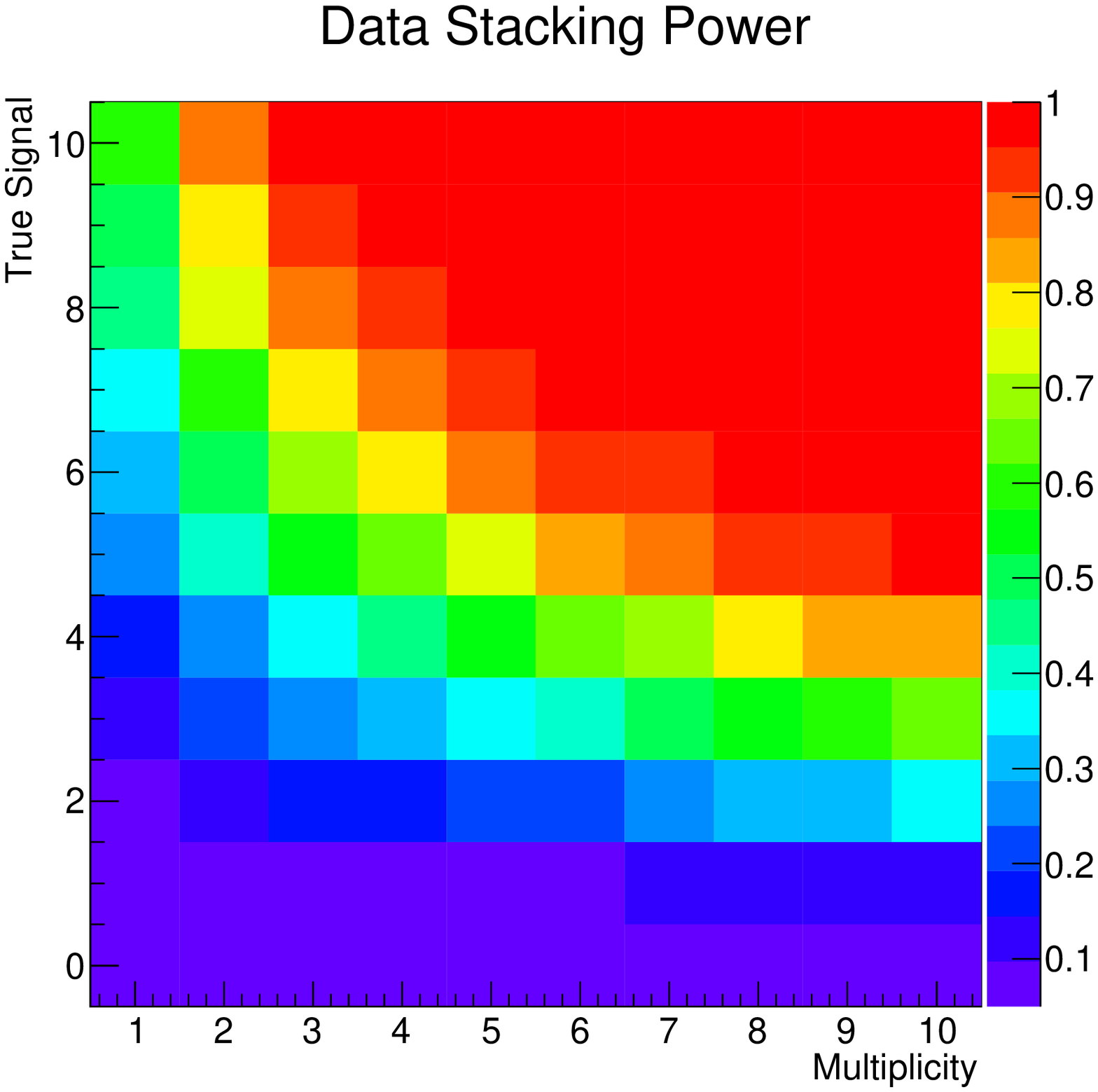}\\
\includegraphics[width=0.45\textwidth]{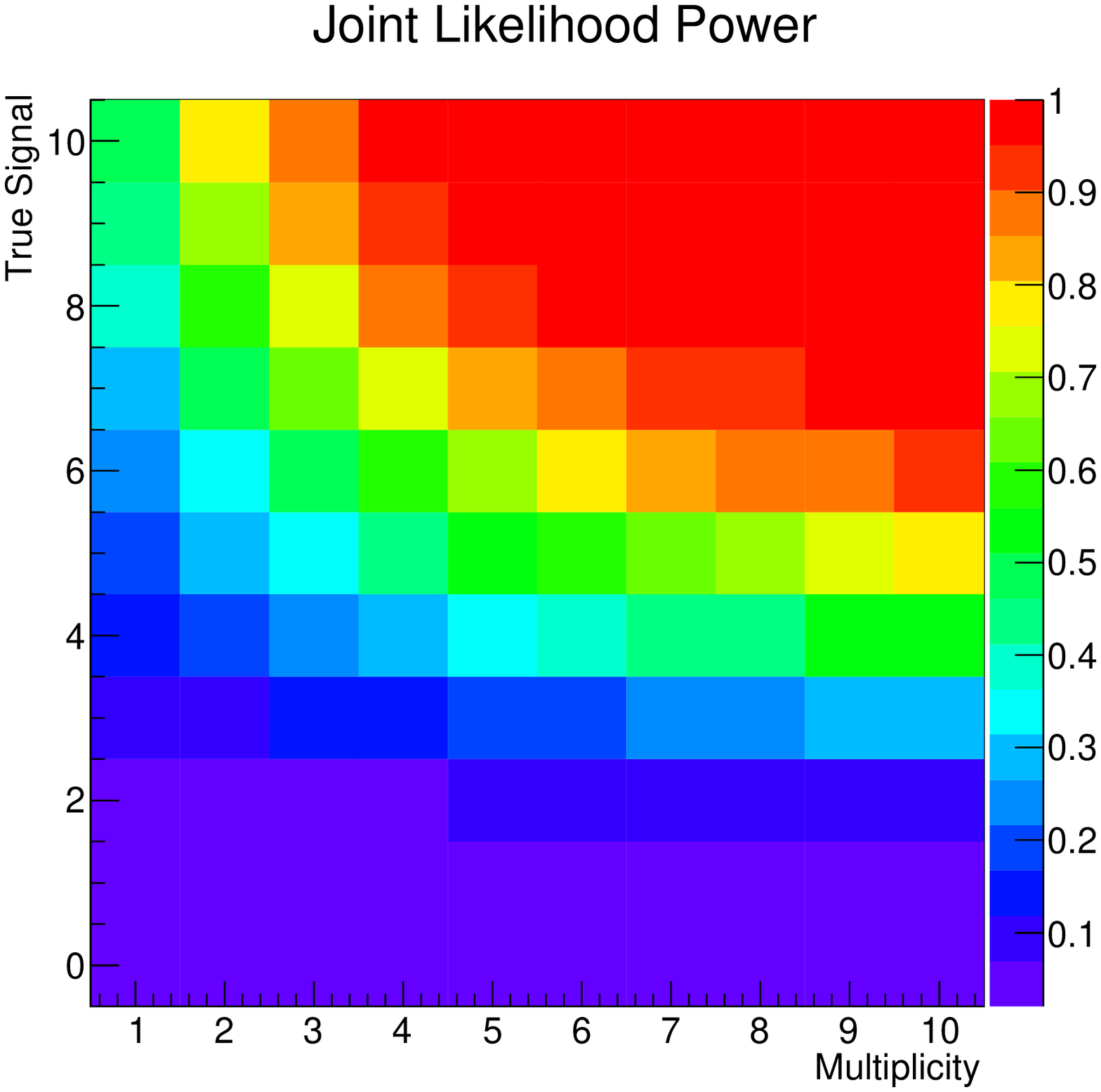}
\includegraphics[width=0.45\textwidth]{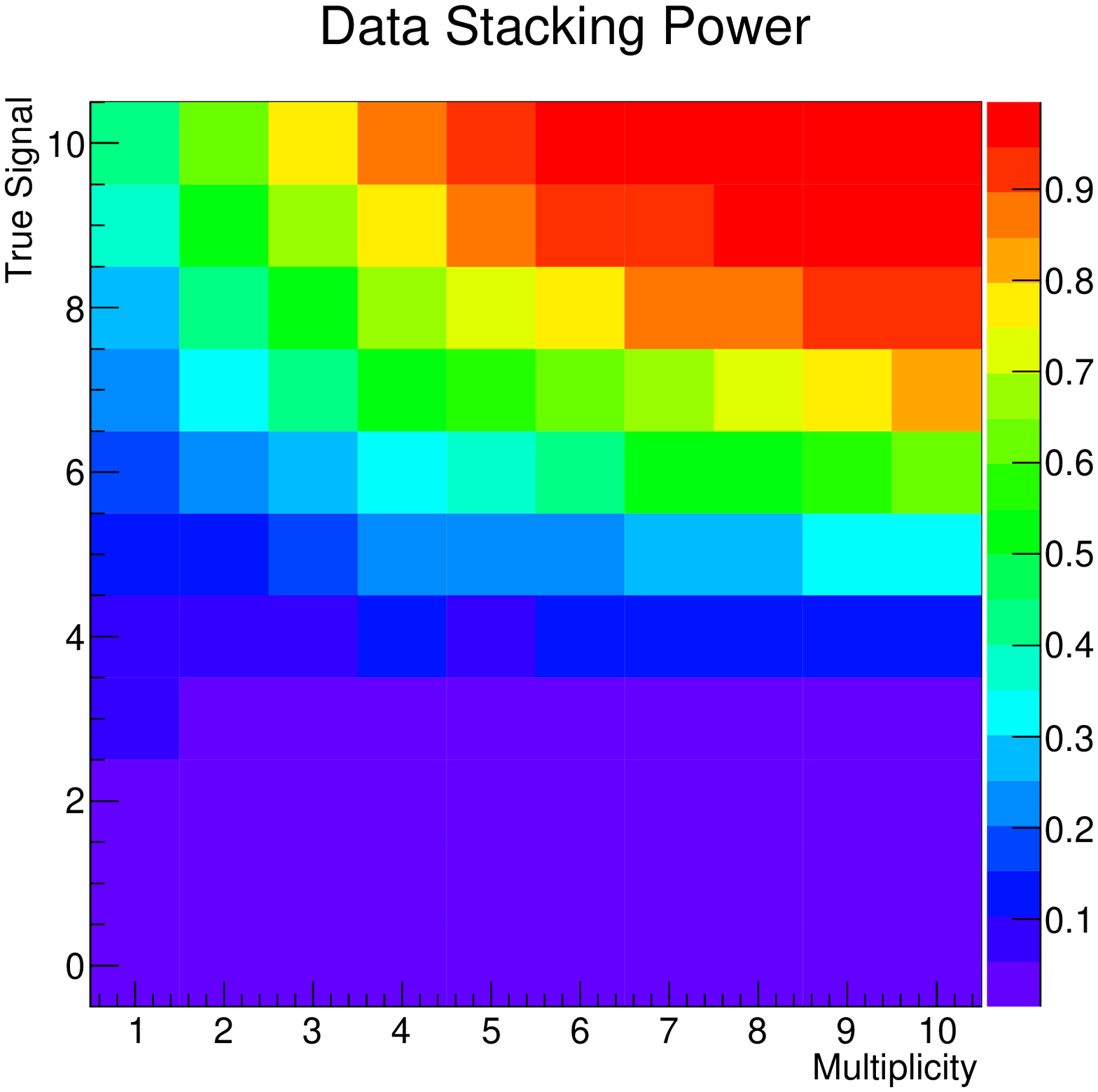}\\
\includegraphics[width=0.45\textwidth]{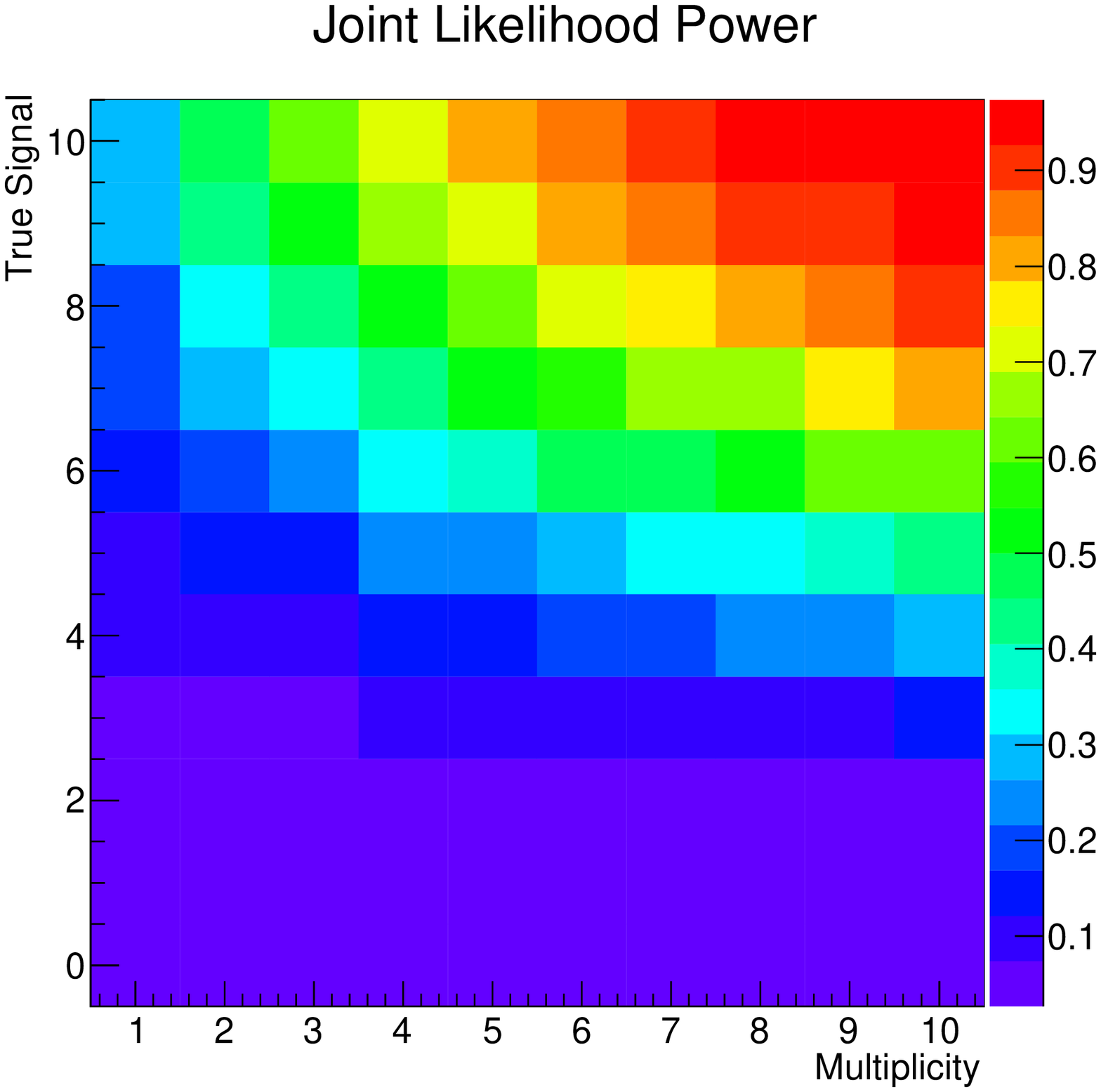}
\includegraphics[width=0.45\textwidth]{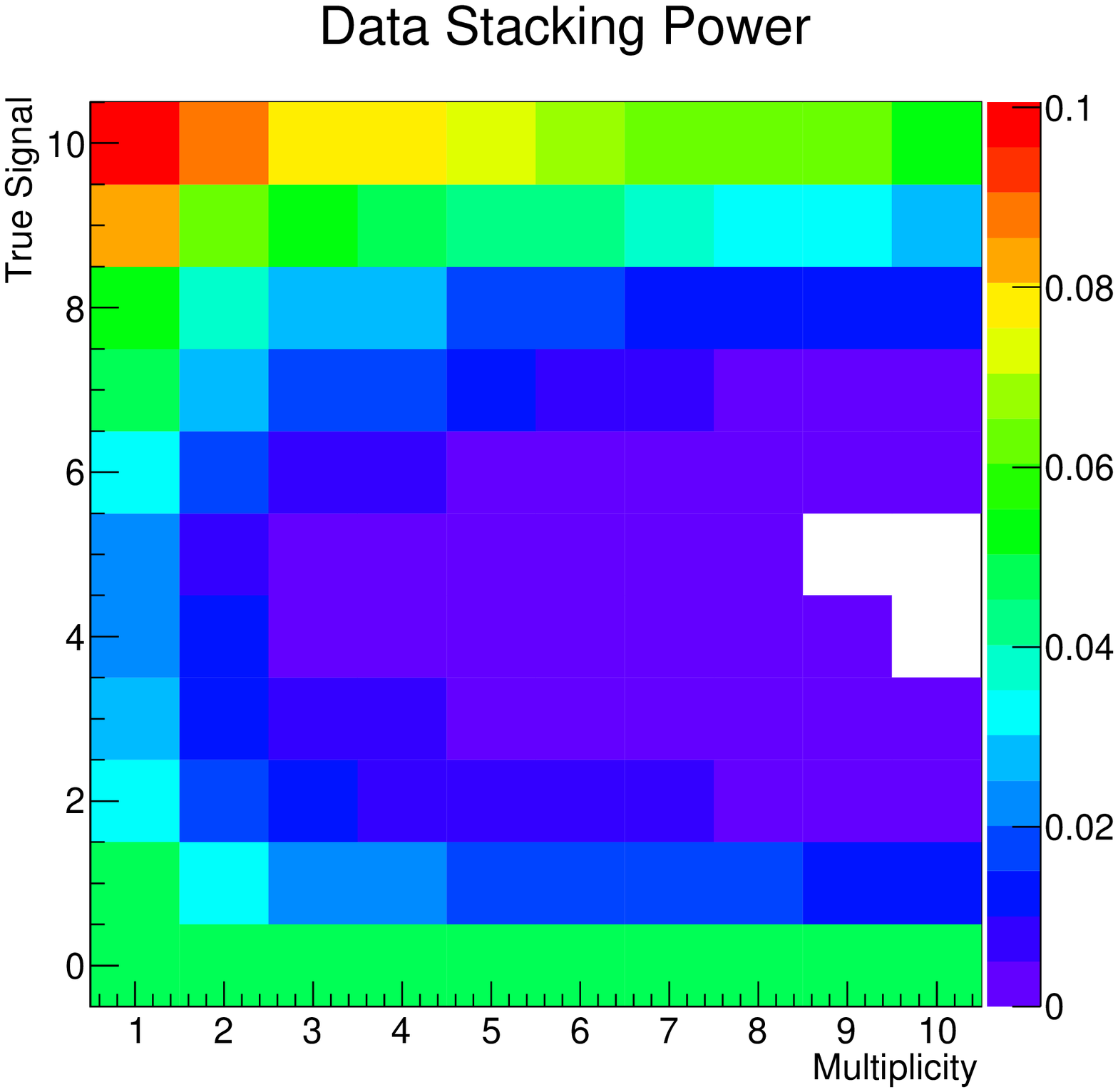}
\end{center}
\caption{Statistical powers derived using $S_{t}$ corresponding to the 95th percentile of the null distributions of $S$ for the Joint Likelihood (\emph{left-hand column}) and Data Stacking (\emph{right-hand column}) approaches. Each bin is derived using analyses of 5000 independent Monte Carlo datasets. The plotted results correspond to Model A (\emph{top row}), Model B (\emph{middle row}) and Model C (\emph{bottom row}) assuming true signals $\tilde{N}_{\rm S}\in[0, 10]$ and target multiplicities $m\in[1,10]$. The underlying datasets are as for Figure \ref{fig:power_diffs}}\label{fig:powers}
\end{figure}

Significance thresholds corresponding to the 95th percentile of the various null distributions of $S$ are plotted in the left-hand column of Figure \ref{fig:power_diffs}. Results derived using the joint-likelihood approach exhibit minimal multiplicity dependence and appear consistent with the nominal $S_{t}=1.96$ expected for a $\chi_{1}$ random variate. Conversely, thresholds which correspond to the data stacking approach systematically exceed their joint likelihood counterparts, and become increasing discrepant for larger $m$ in simulations that assume asymmetric distributions of $\alpha$. 
Accordingly, recourse to a nominal distribution of $S$ is not appropriate in the data stacking framework, and valid interpretation of the significance of a particular detection requires empirical calibration of distinct significance thresholds for each target multiplicity. 
The values of $S_{t}$ plotted in Figure \ref{fig:power_diffs} permit derivation of appropriately calibrated estimates of the statistical power at each vertex of the investigated model parameter space. The resultant probabilities are presented in Figure \ref{fig:powers}, while the results plotted in the right-hand column of Figure \ref{fig:power_diffs} reveal the corresponding differences power between the alternative stacking techniques.

Results pertaining to Models A and B exhibit saturation at extreme values of $\tilde{N}_{\rm S}$ and $m$, caused by the limited statistics of the Monte Carlo datasets. Results corresponding a symmetric Gaussian distribution for $\alpha$ reveal derived powers which are compatible to $<2\%$ for all $m\in[1,10]$ and $\tilde{N}_{\rm S}\in[0, 10]$. Although traditional data stacking appears to offer marginally superior performance for this scenario, asymmetric parameterisations for the distribution of $\alpha$ reverse this outcome. Indeed, powers derived using the joint likelihood approach in conjunction with Models B and C exhibit substantial enhancement with respect to their data stacking counterparts. Moreover, powers corresponding to the data stacking technique exhibit a marked sensitivity to the degree of asymmetry in the adopted $\alpha$ distribution, which appears to be substantially ameliorated by the joint likelihood approach. Strong suppression of the resultant data stacking power is evident when Model C is applied to the Monte Carlo analysis, with derived discrepancies $\sim0.9$ separating the alternative stacking techniques at the upper extremes of the modelled parameter ranges. More significantly for practical applications, assuming the intermediate scenario represented by Model B, reveals a maximum disparity of $\sim 0.4$ for the experimentally interesting region of $\tilde{N}_{\rm S}\sim5$ as $m\rightarrow10$. In this context, the joint likelihood approach yields a two-fold enhancement in the likelihood of detecting a genuine $\gamma$-ray signal.

\subsection{Confidence Intervals}

For an analysis to perform reliably in an experimental context, it is important that any generated confidence intervals have correct coverage. In reality, despite sophisticated treatment of associated systematic and statistical uncertainties, residual disparities between the assumed and actual distributions of experimental data often prevent practical realisation of this criterion. Moreover, deviations from correct coverage often lead to an unexpected increase in the frequency of spuriously inferred scientific conclusions. Accordingly, the degree by which the derived confidence intervals deviate from their nominal coverage provides an informative comparator for the alternative stacking techniques.

\begin{figure}[h]
\begin{center}
\includegraphics[width=0.45\textwidth]{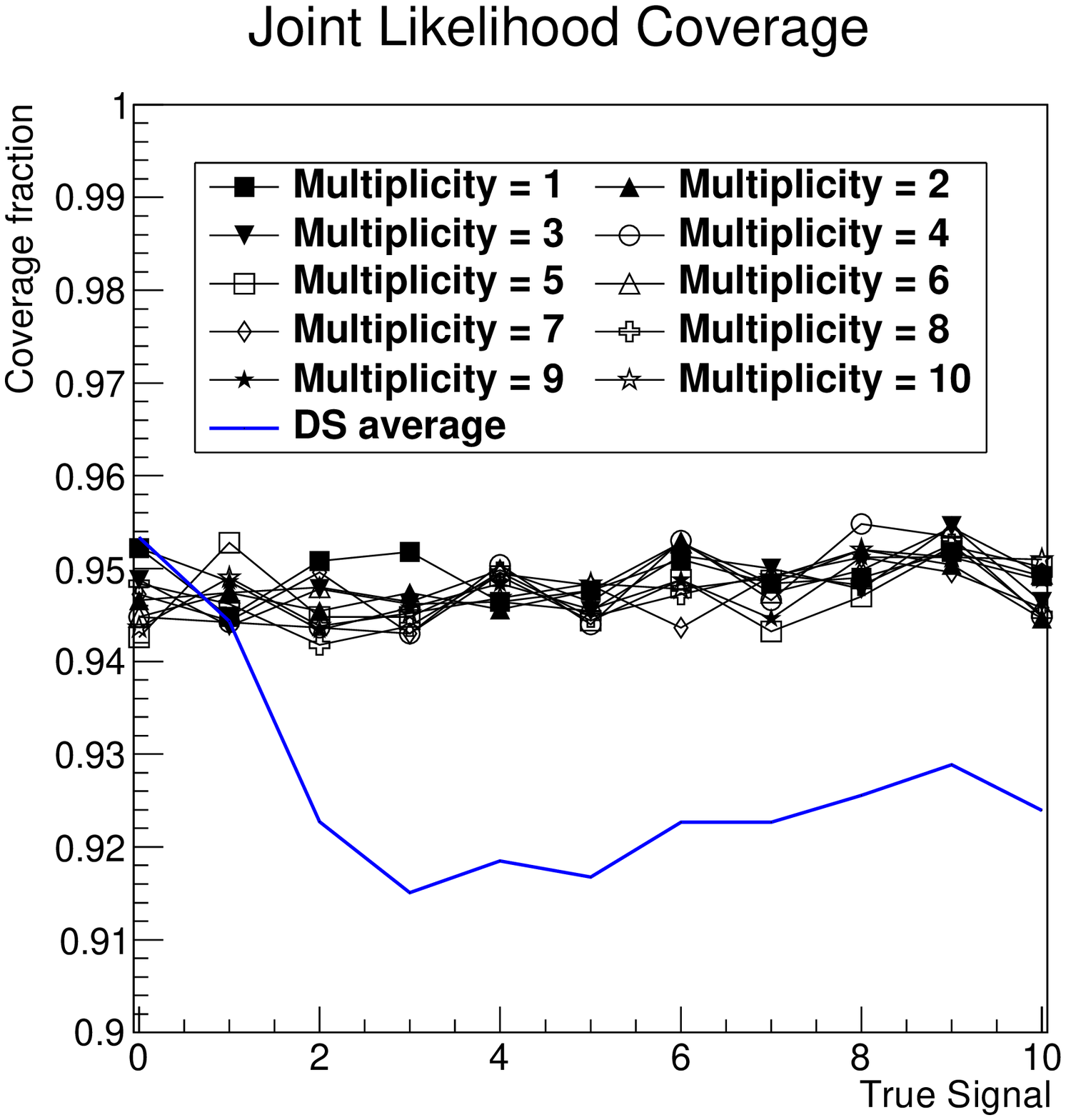}
\includegraphics[width=0.45\textwidth]{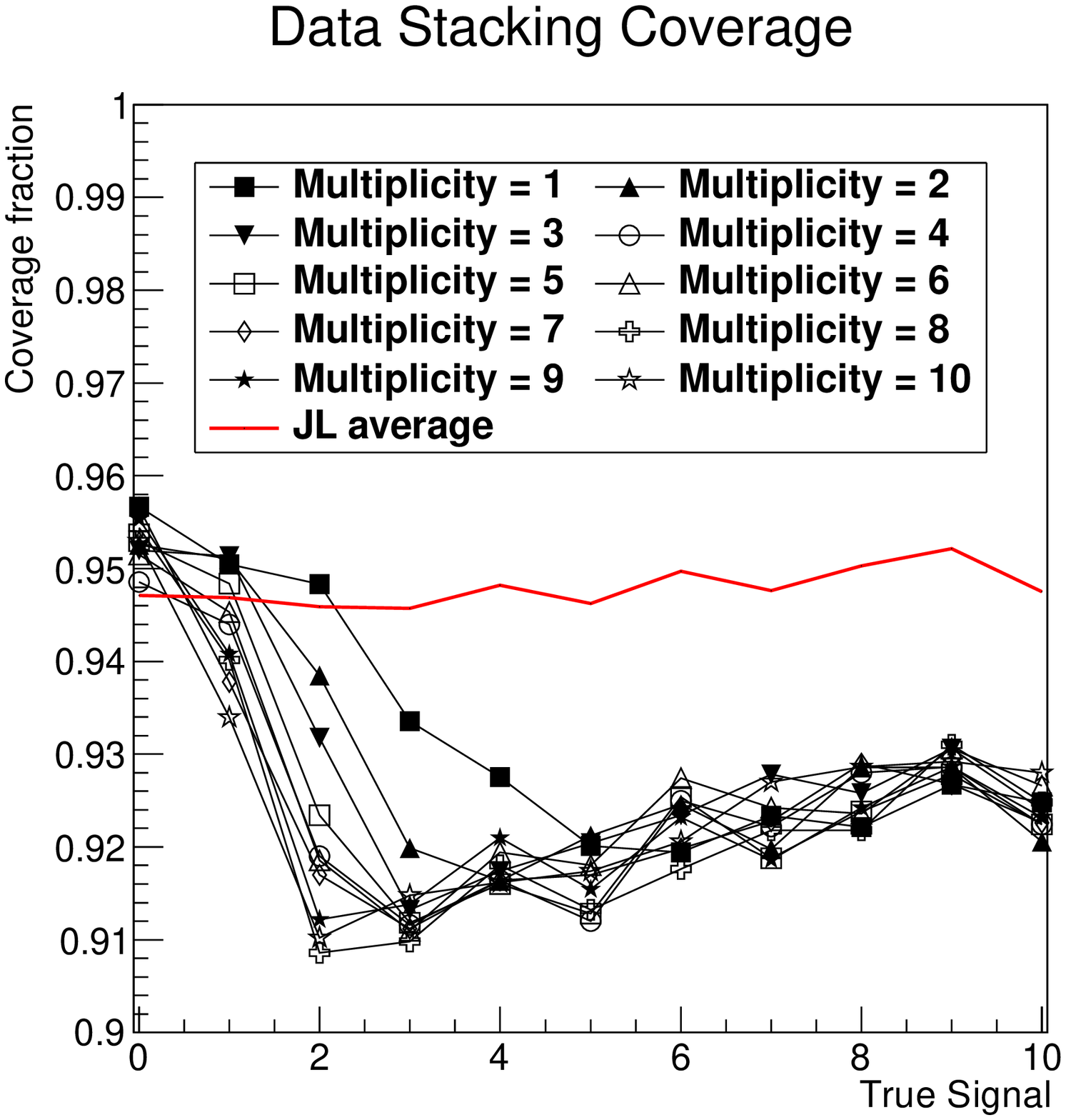}\\
\includegraphics[width=0.45\textwidth]{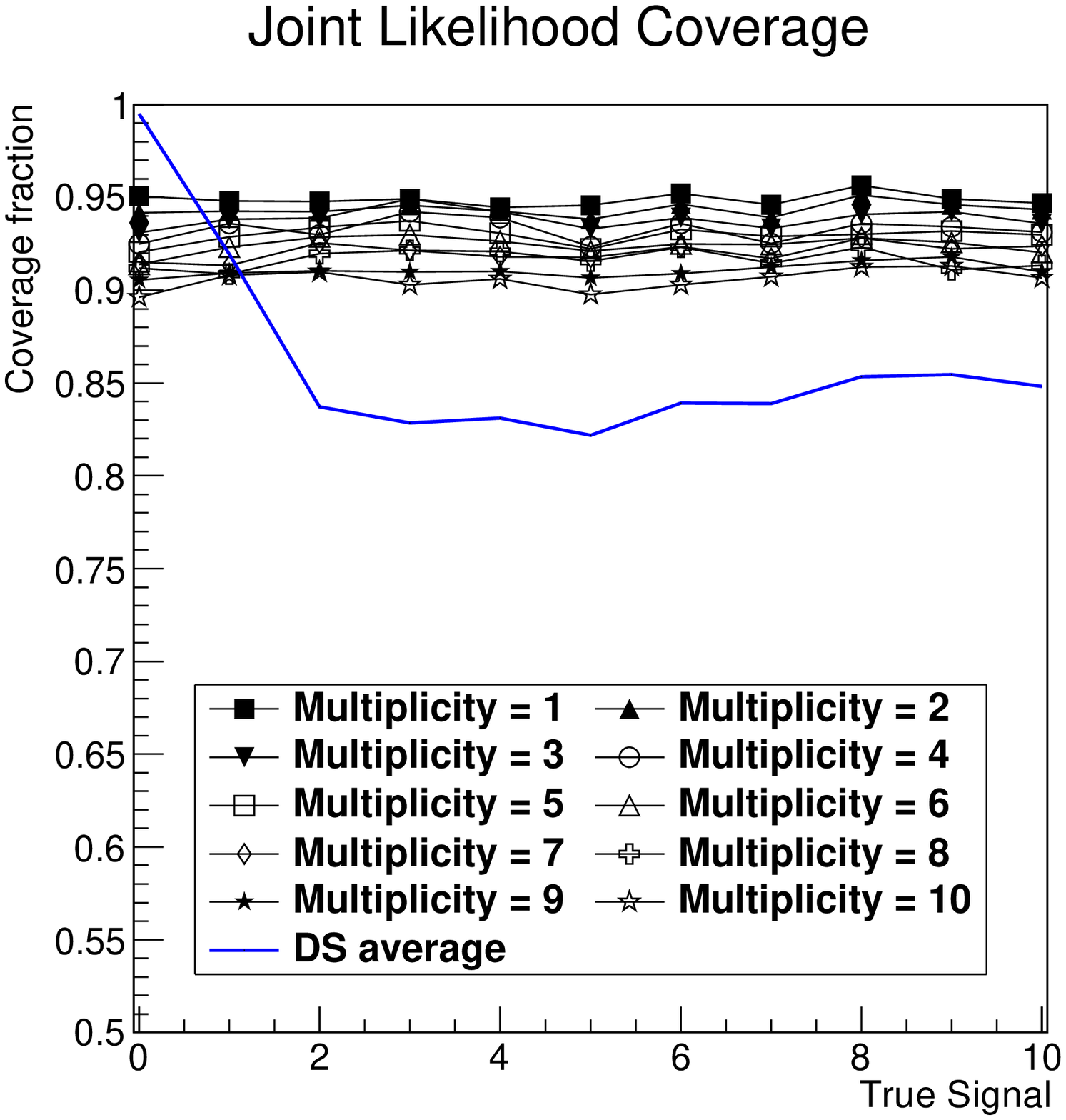}
\includegraphics[width=0.45\textwidth]{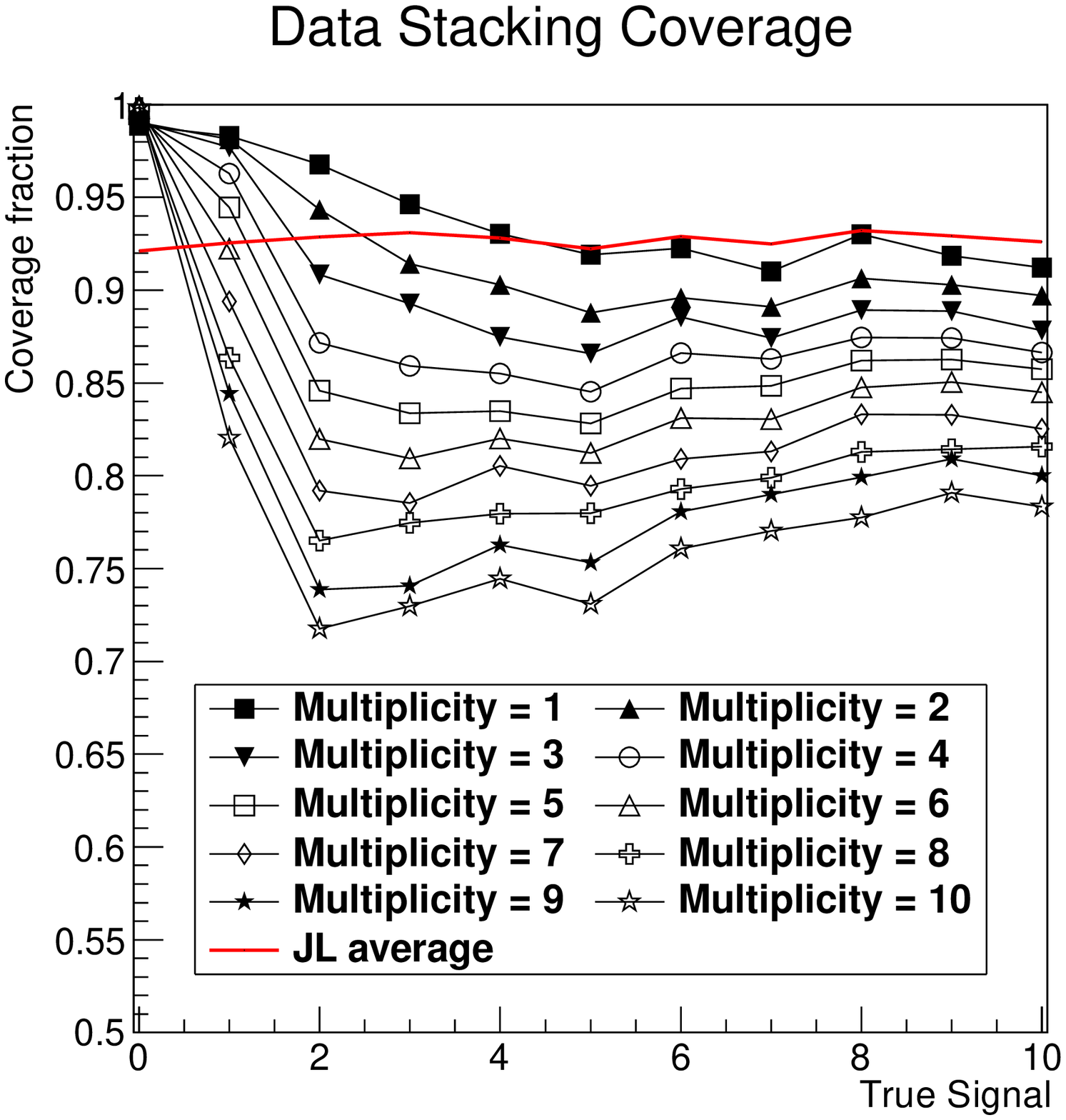}\\
\includegraphics[width=0.45\textwidth]{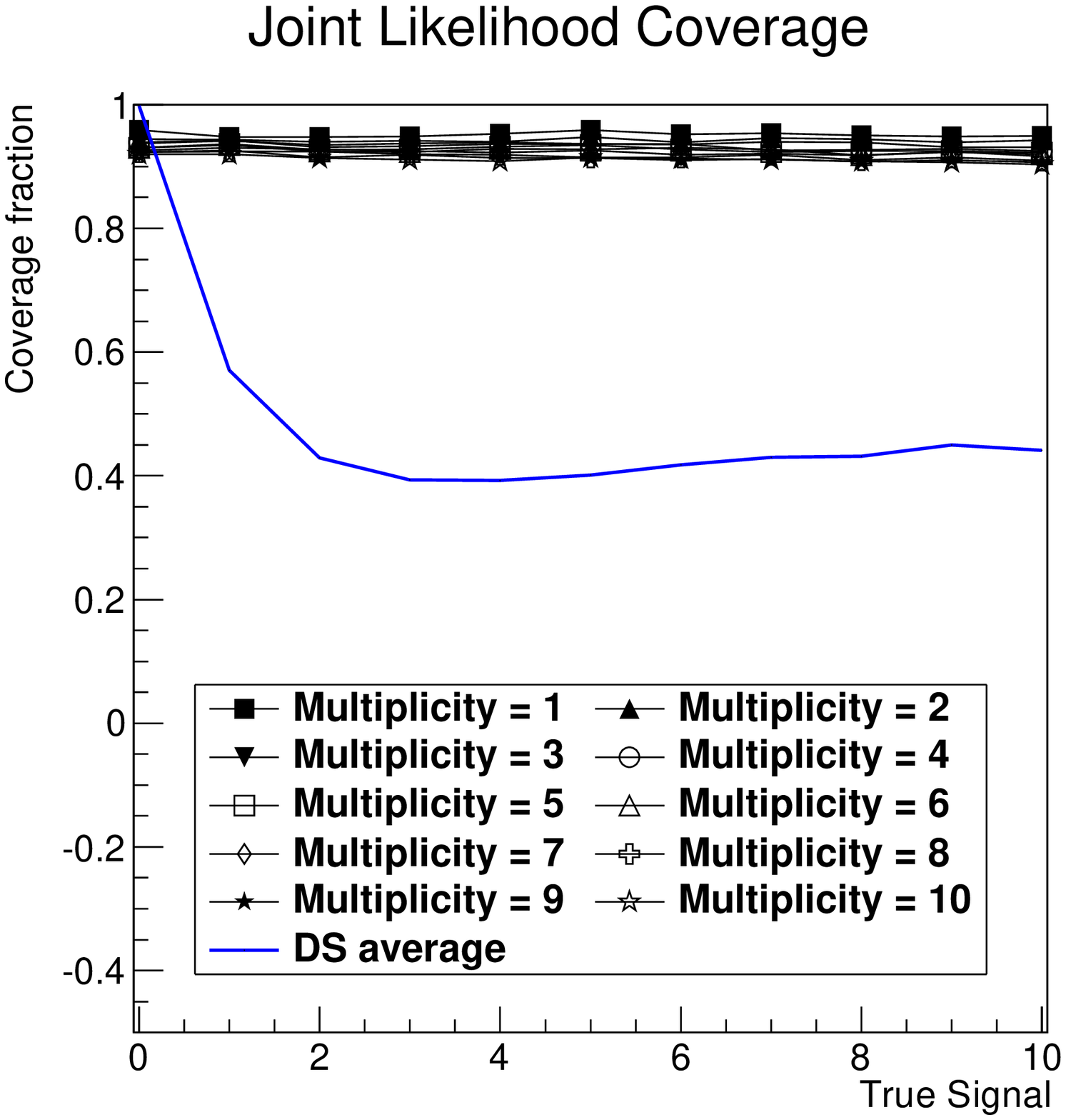}
\includegraphics[width=0.45\textwidth]{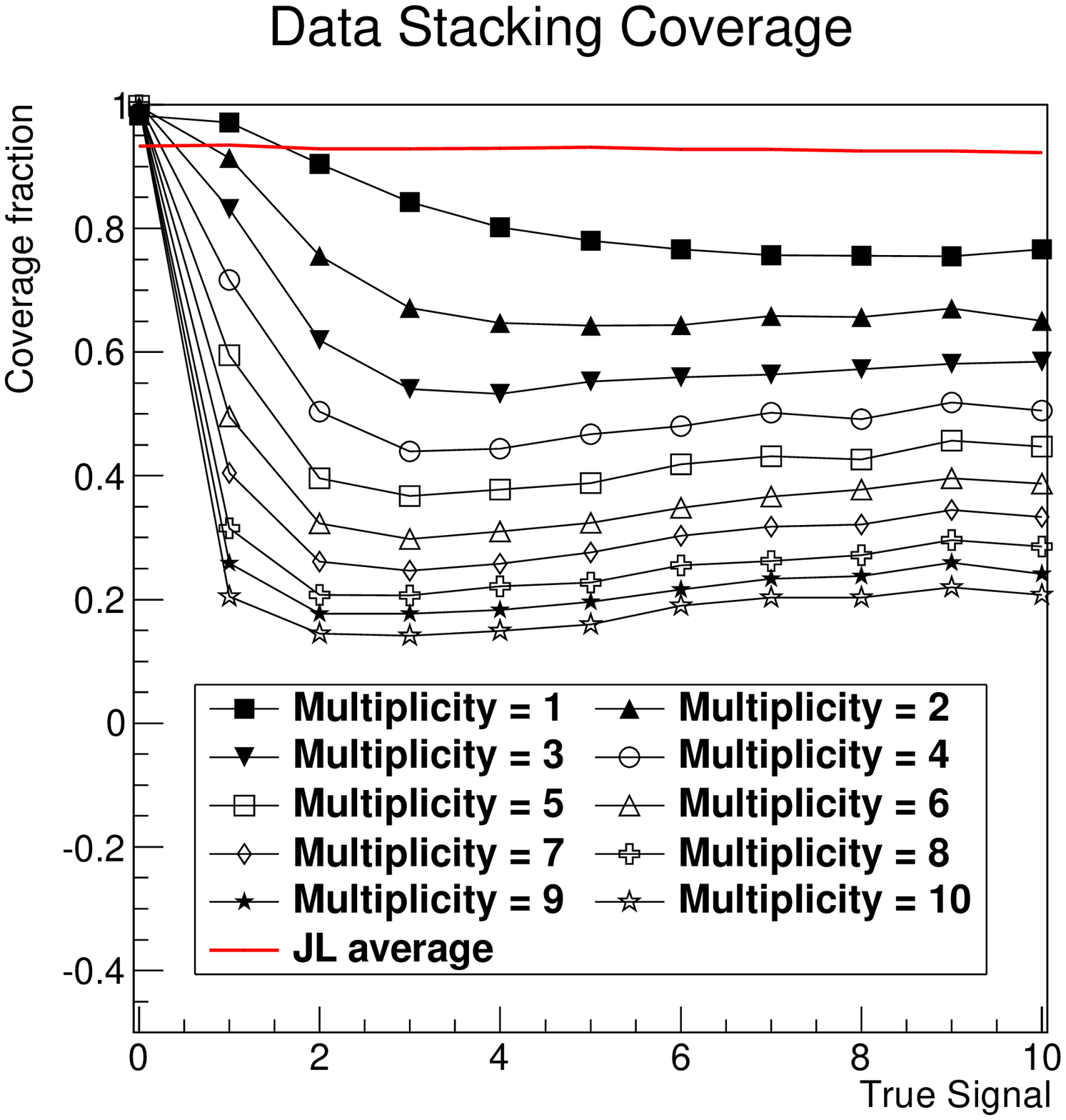}
\end{center}
\caption{Joint Likelihood (JL) and Data Stacking (DS) coverage characteristics corresponding Model A (\emph{top row}), Model B (\emph{middle row}) and Model C (\emph{bottom row}). Each datapoint is derived using analyses of 5000 independent Monte Carlo datasets and indicates the fraction of generated 95\% confidence intervals on $N_{\rm S}$ which include the true signal value, $\tilde{N}_{\rm S}$. The underlying datasets are as for Figure \ref{fig:power_diffs}.}\label{fig:coverage_fractions}
\end{figure}
\begin{figure}[h]
\begin{center}
\includegraphics[width=0.45\textwidth]{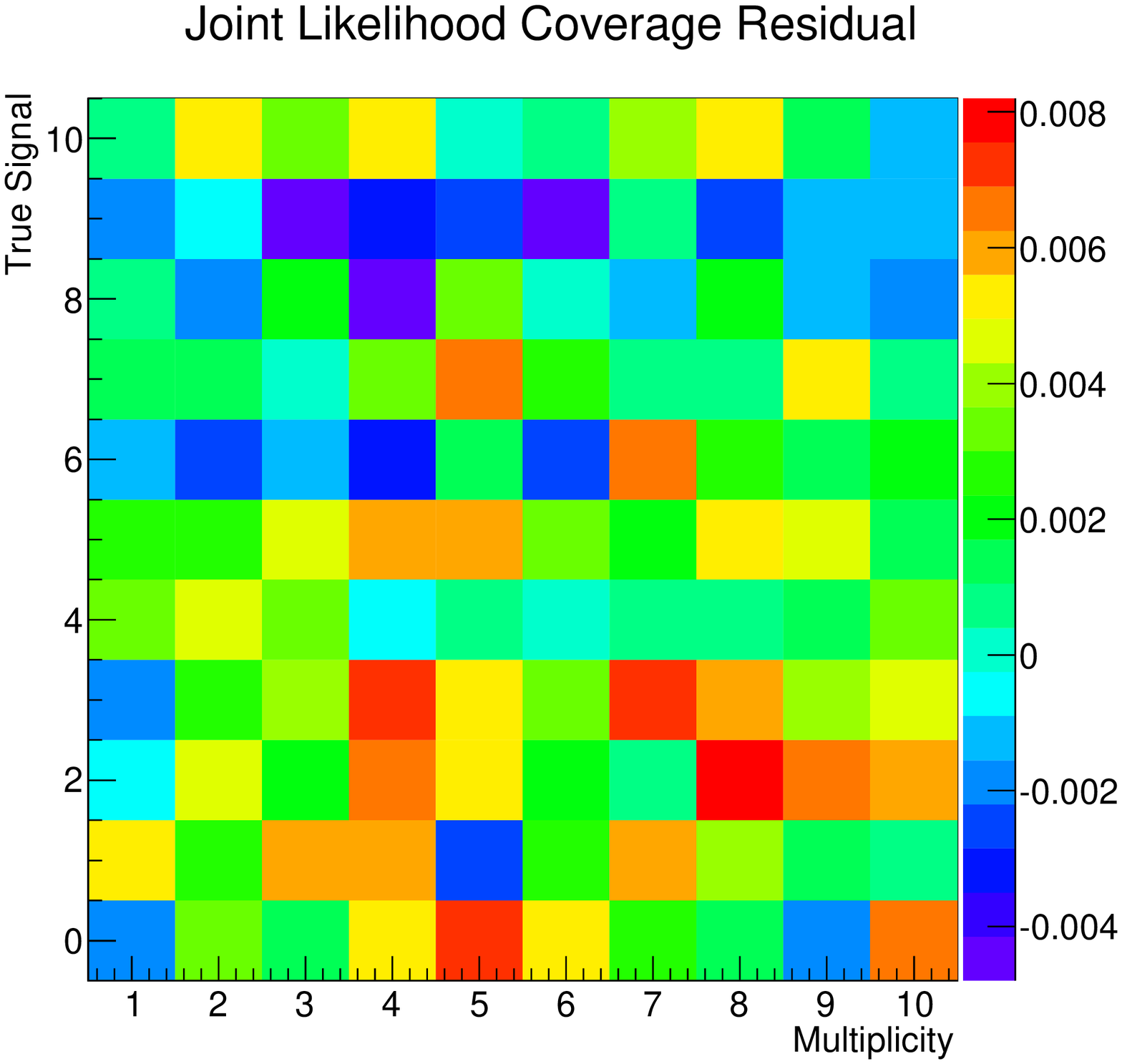}
\includegraphics[width=0.45\textwidth]{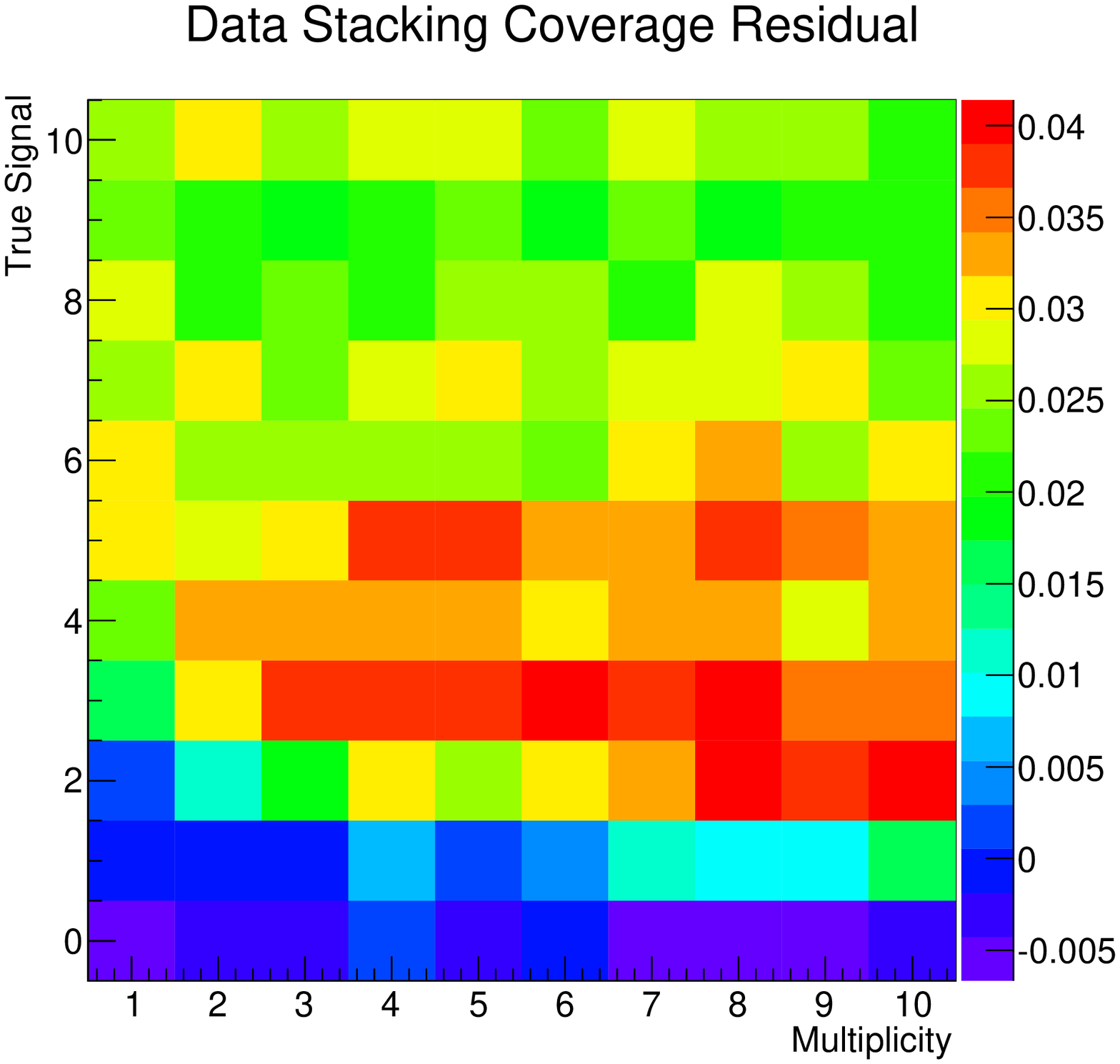}\\
\includegraphics[width=0.45\textwidth]{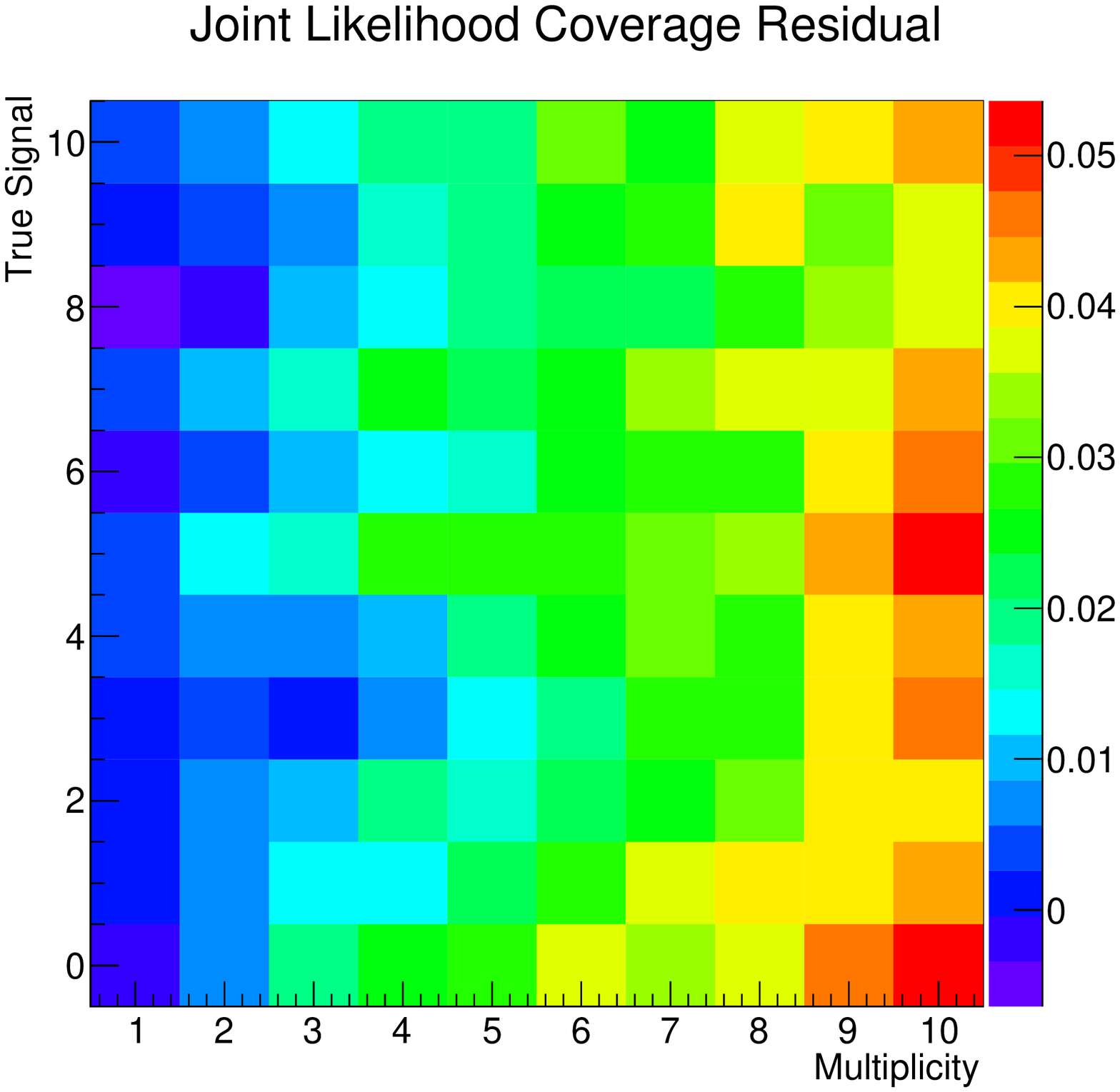}
\includegraphics[width=0.45\textwidth]{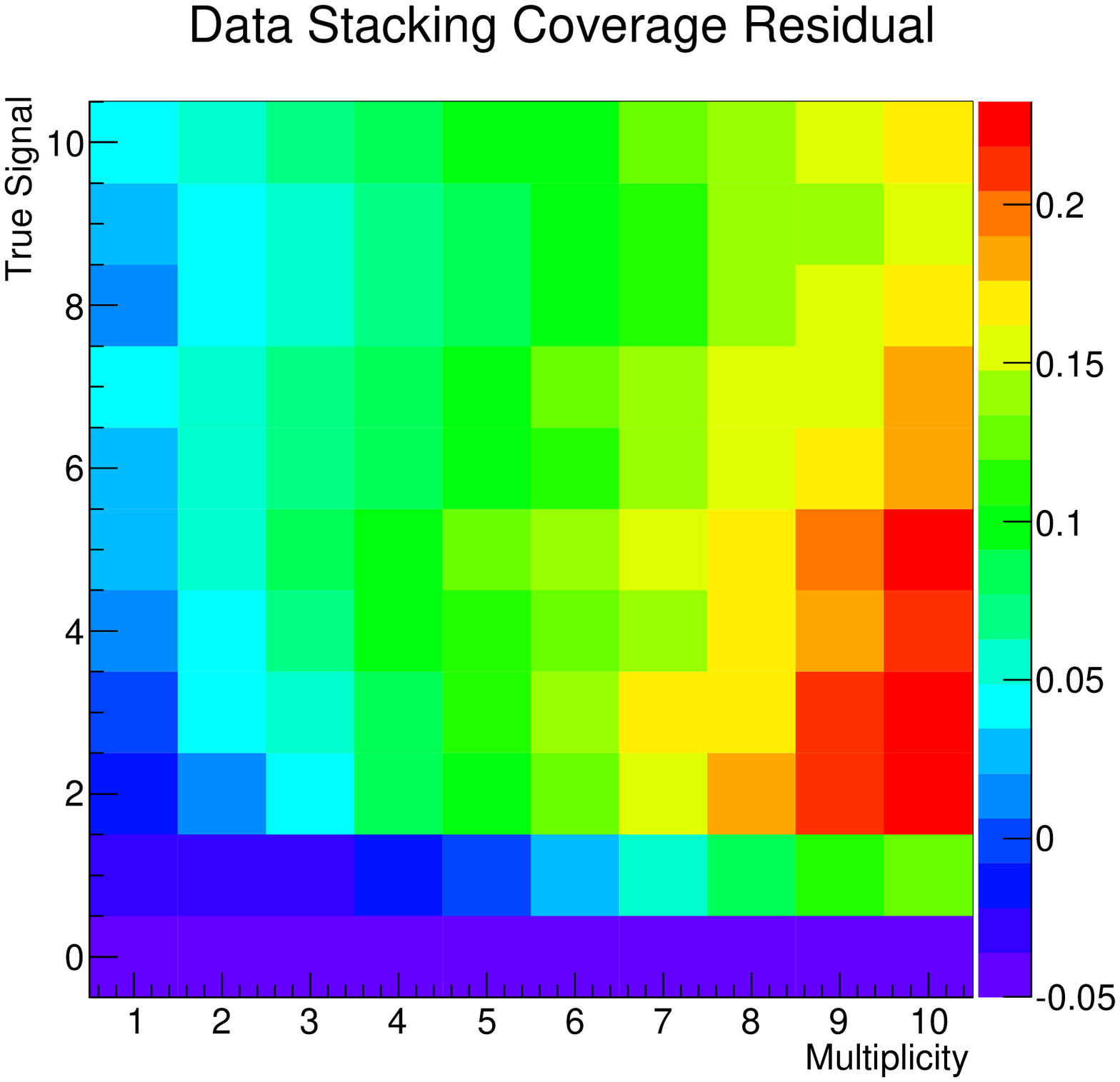}\\
\includegraphics[width=0.45\textwidth]{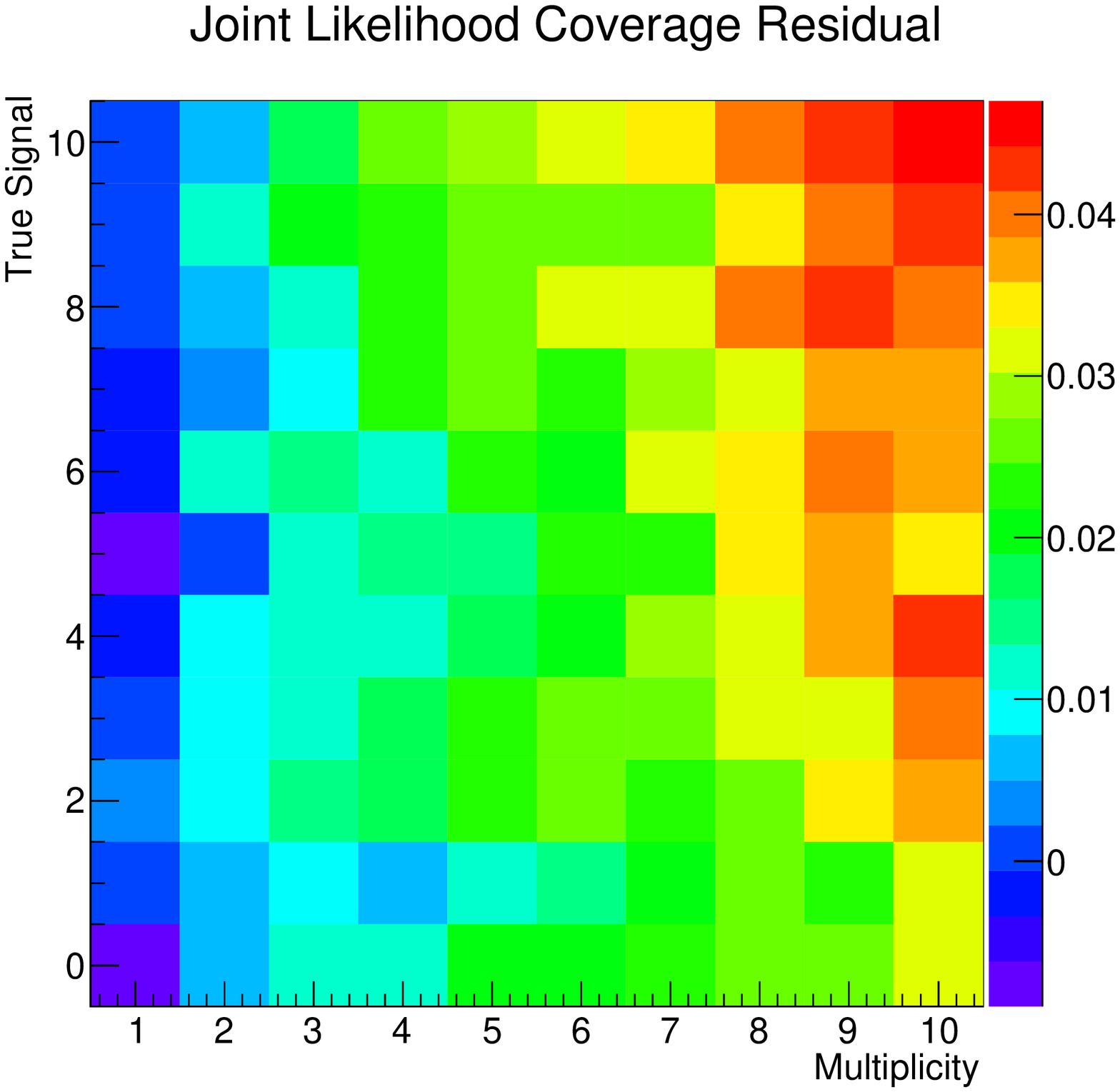}
\includegraphics[width=0.45\textwidth]{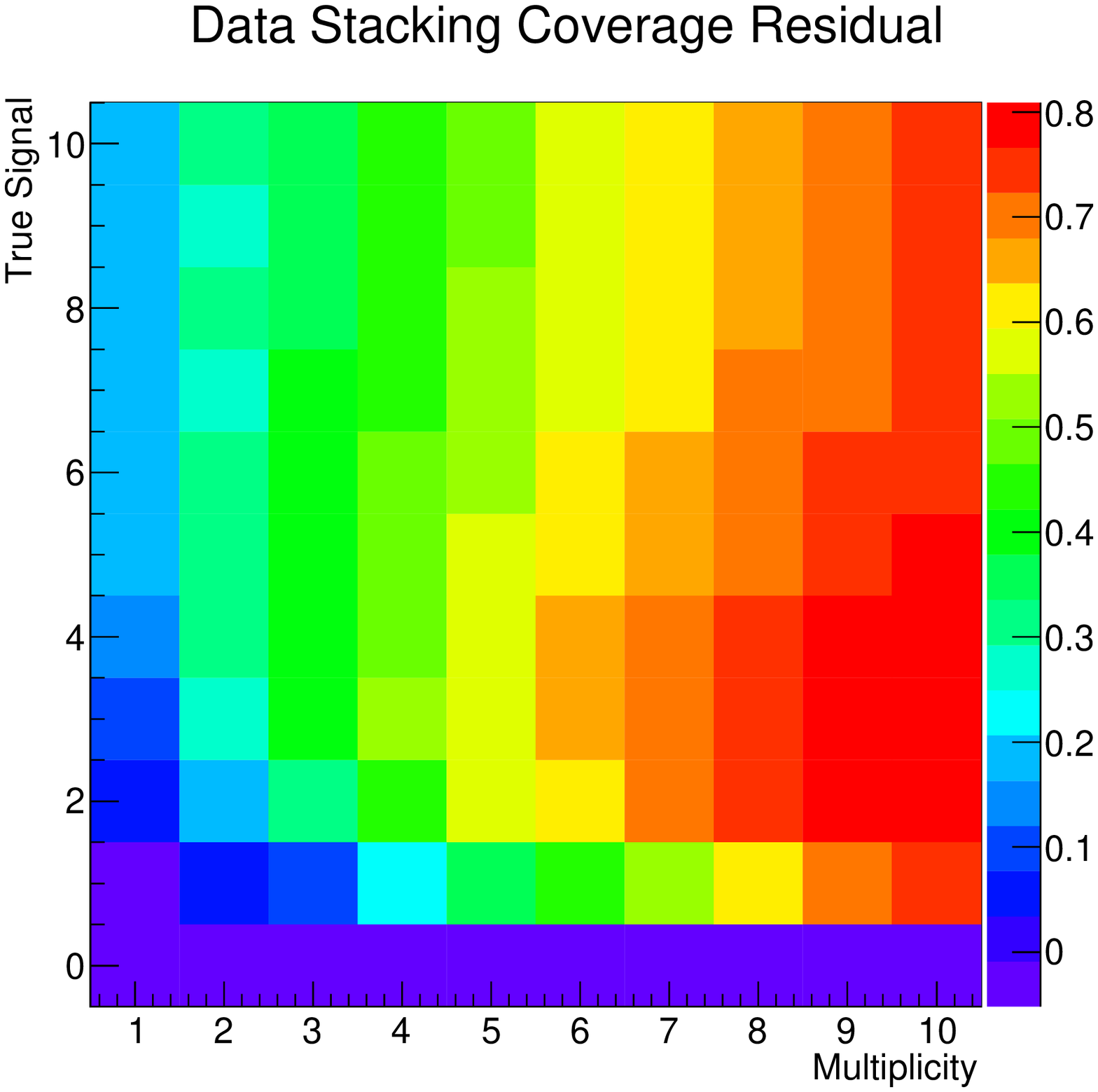}
\end{center}
\caption{Residuals with respect to nominal coverage for 95\% confidence intervals on $N_{\rm S}$ corresponding to the Joint Likelihood (JL) and Data Stacking (DS) approaches. Results are shown which correspond Model A (\emph{top row}), Model B (\emph{middle row}) and Model C (\emph{bottom row}). The underlying datasets are as for Figure \ref{fig:power_diffs}.}\label{fig:coverage_residuals}
\end{figure}

Figure \ref{fig:coverage_fractions} displays the fraction of derived 95\% confidence intervals for $N_{\rm S}$ which cover the true signal value, for each parameterisation of $G(\alpha|\tilde{\alpha})$. Complementary insight is provided by Figure \ref{fig:coverage_residuals}, which illustrates the corresponding deviations from correct coverage as a function of $\tilde{N}_{\rm S}$ and the target multiplicity, $m$. It is evident that confidence intervals generated by the traditional data stacking approach are generally incompatible with their stated coverage. While the derived coverage fractions exceed the nominal 95\% for low signals and target multiplicities, the majority of the $(\tilde{N}_{\rm S},m)$ parameter space is characterised significant negative deviations. This behaviour is typically described as \emph{under-coverage}, and implies excessively permissive confidence intervals with an increased propensity to generate false-positive results. The observed under-coverage is maximised for small non-zero signals and appears to be exacerbated by the addition of targets to the ensemble dataset. Moreover, increased asymmetry of the modelled $\alpha$ distribution has a markedly detrimental effect on the derived coverage characteristics. Indeed, the maximum divergence from correct coverage increases from 4\% to 25\% and then to 85\% for Model A, Model B, and Model C, respectively. In the worst case, this means that only 10\% of the derived intervals include the true parameter value.
The inherent inability of the data stacking technique to acknowledge systematic uncertainties leads to confidence intervals which are unrealistically narrow. Consequently, reported accuracy of experimental measurements will be excessively optimistic and in the worst cases could be used to infer spurious scientific conclusions.

In contrast, notable stability in the coverage fractions derived using the joint likelihood technique indicates robust handling of systematic uncertainties. Confidence intervals derived under the assumption of a Gaussian distribution in $\alpha$ exhibit coverage which is correct to within $\lesssim1\%$ throughout the studied region of the model parameter space. Variation of the coverage fraction as a function of $\tilde{N}_{\rm S}$ remains minimal when asymmetric parametrisations of $G(\alpha|\tilde{\alpha})$ are assumed. Although under-coverage which worsens with increasing target multiplicity is evident in results that correspond to Models B and C, the discrepancy with respect to correct coverage does not exceed $\sim5\%$.

\section{Discussion}\label{sec:discussion}

The comparative evaluation presented in $\S$\ref{sec:results} reveals several limitations to the applicability of traditional data stacking in situations when $\alpha$ is uncertain, many of which are addressed by the joint likelihood technique.

Reliable inference of signal characteristics is impaired by significant levels of under-coverage for $\tilde{N}_{\rm S}\gtrsim2$ in the data stacking approach. Moreover, the inherent permissivity of derived confidence intervals is compounded by the inclusion of additional targets. In combination, these characteristics imply a substantially increased tendency to generate spurious results which is exacerbated as the stacked target ensemble is enlarged. In an experimental context, this behaviour would be highly undesirable and largely undermines the motivation for combined source analysis. In contrast, confidence intervals provided by the joint likelihood approach closely approximate correct coverage under the assumption of symmetric uncertainties on $\alpha$, and exhibit multiplicity-dependent under-coverage at $\lesssim5\%$ levels if an asymmetric distribution is adopted. 

The deviations from correct coverage that are exhibited by the data stacking approach are consistent with the erroneous assumption of negligible uncertainty in the value of $\alpha$. Resolution of this issue is complicated by the lack of a straightforward prescription for propagating systematic uncertainties to the calculated value of $\bar{\alpha}$. Conversely, the joint likelihood method provides a robust technique for the treatment of uncertain $\alpha$ estimates on a target-specific basis. 

Appropriate interpretation of the combined signal significance is also more complicated within the data stacking framework. The Monte Carlo null distributions for $S$ presented in Figure \ref{fig:null_sigma_dists} indicate that application of Equation \eqref{eq:LiMa17} requires proper calibration of $S_{t}$ to prevent spurious rejection of the null hypothesis. The situation deteriorates when increasingly asymmetric parameterisations of the generated $\alpha$ distribution are adopted. Under these circumstances, na\"\i ve assumption of a nominal $\chi_{1}$ distribution for $S$ induces a systematic overestimation of corresponding significance levels which worsens with increasing target multiplicity. In an experimental context, the implied increase in the false-positive detection rate for larger target samples is contrary to common expectation and therefore represents a serious shortcoming of the data stacking technique.
In contrast, significance estimates derived for $\tilde{N}_{\rm S}$ using the joint likelihood approach appear insensitive to the value of $m$ and display excellent correspondence with the expected $\chi_{1}$ distribution.

Notwithstanding appropriate calibration of a suitable detection threshold $S_{t}$, the powers which characterise the data stacking technique indicate that asymmetry in the adopted $\alpha$ distribution substantially diminishes the overall sensitivity. Conversely, the powers corresponding to the joint likelihood analysis appear relatively unaffected by the choice of $G(\alpha|\tilde{\alpha})$. The alternative approaches exhibit similar properties, and indeed data stacking appears to marginally outperform the joint likelihood approach, if symmetric systematic uncertainties are assumed. 

\begin{figure}[h]
\begin{center}
\includegraphics[width=0.48\textwidth]{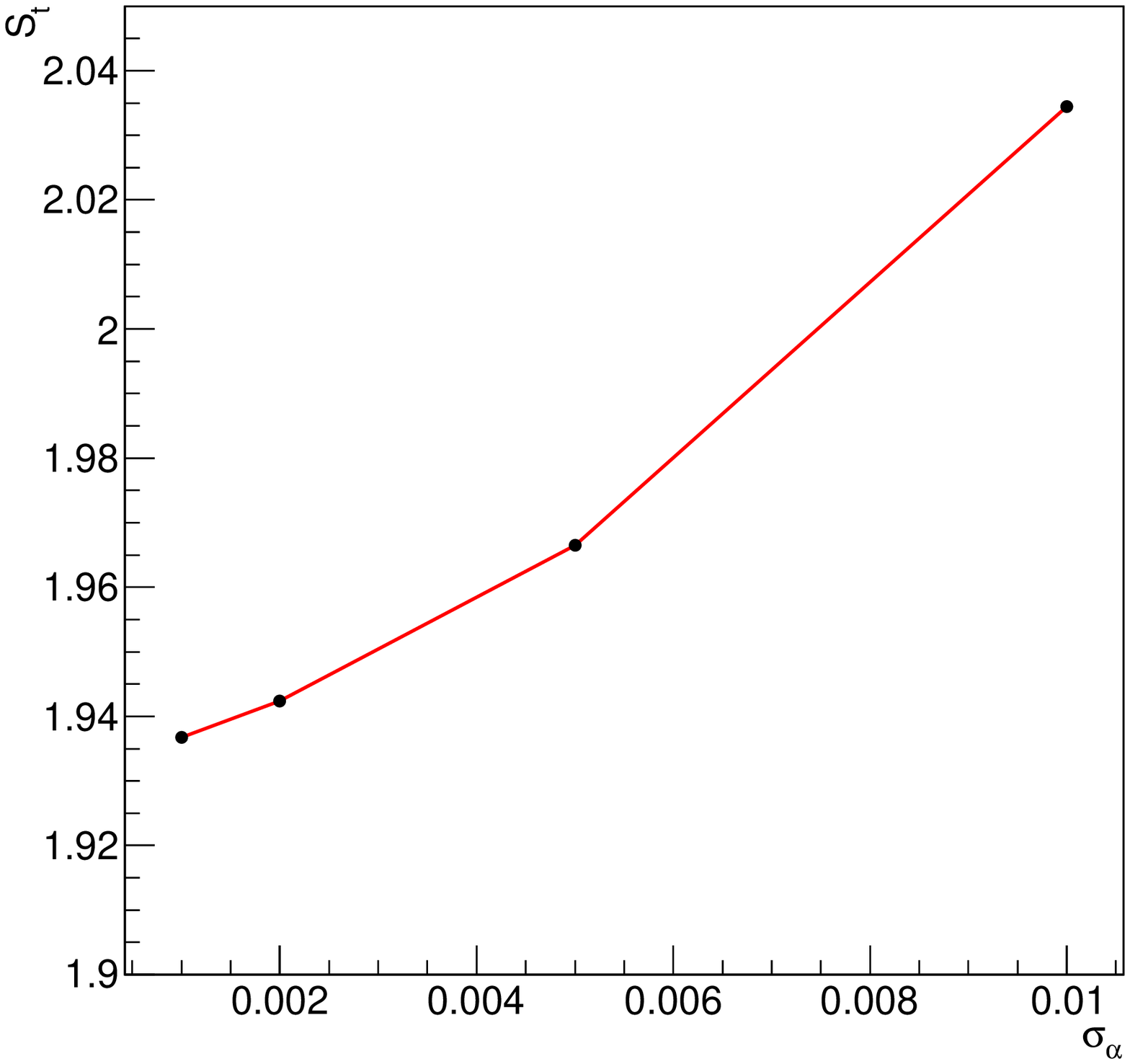}
\includegraphics[width=0.48\textwidth]{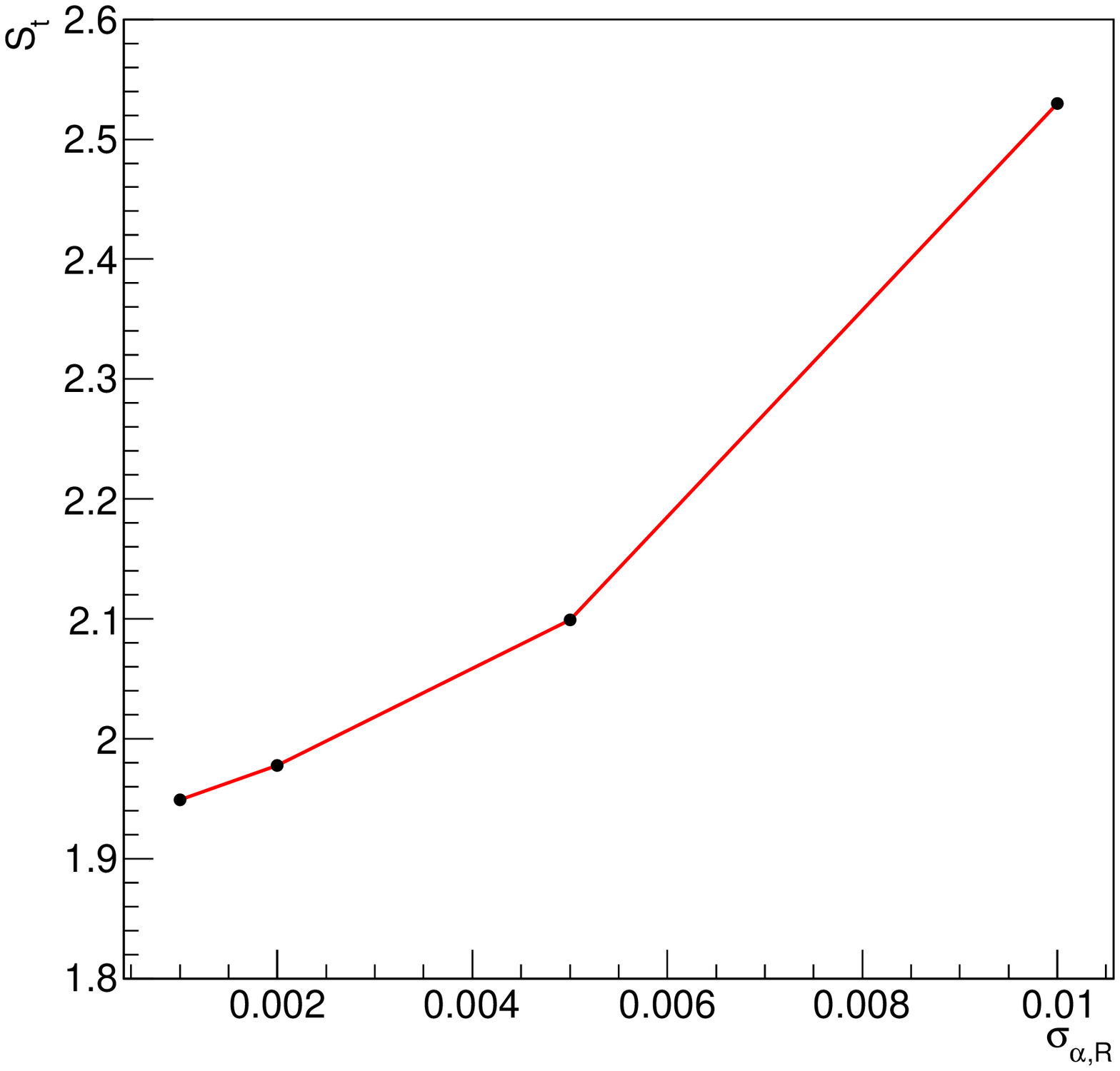}
\end{center}
\caption{Calibrated significance threshold values corresponding to the 95th percentile of the Monte Carlo null distributions of $S$ for the and Data Stacking approach. The illustrated results correspond to Gaussian (\emph{left}) and bifurcated Gaussian (\emph{right}) parameterisations for $G(\alpha|\tilde{\alpha})$, using target-specific uncertainties $1\%\leq\sigma_{\alpha}/\tilde{\alpha} \leq 10\%$ for the symmetric models, and $1\%\leq\sigma_{\alpha, \rm{R}}/\tilde{\alpha}  \leq 10\%$ with $\sigma_{\alpha, \rm{R}} = 2\sigma_{\alpha, \rm{L}}$ for the asymmetric models. Each datapoint is derived using analyses of 5000 independent Monte Carlo datasets, which assume ensembles of $m=50$ targets having true signals $\tilde{N}_{\rm S}= 0$. All Monte Carlo datasets were generated assuming $\tilde{N}_{\rm OFF} = 100$ and $\tilde{\alpha} = 0.1$.}\label{fig:delta_alpha_mag}
\end{figure}

It should be acknowledged that the degree of adverse behaviour inherent in the data stacking technique is contingent upon the adopted values of $\tilde{\alpha}$ and $\tilde{N}_{\rm OFF}$, as well as the assumed magnitude of the simulated systematic effects. 
Accurate modelling of the system acceptance may ameliorate the observed pathologies, with the cumulative effect of target-specific $\alpha$ uncertainties only becoming significant for target ensembles with $m\gg10$. Figure \ref{fig:delta_alpha_mag} illustrates the combined effect on data stacking of systematic offsets at the few percent level using calibrated values of the significance threshold that correspond to the 95th percentile of simulated null distributions for $S$. The Monte Carlo data sets assume relative target-wise uncertainties $0.01\leq\sigma_{\alpha,i}/\tilde{\alpha} \leq 0.1$, for both symmetric and asymmetric parameterisations for $G(\alpha|\tilde{\alpha})$. The curves correspond to simulated ensembles comprising $m=50$ target sources for which $\tilde{\alpha}=0.1$ and $\tilde{N}_{\rm OFF}=100$. Despite an inevitable reduction in the influence of the modelled systematic offsets, it is clear that even small target-specific biasses can reinforce to distort the combined significance for large observational datasets. Moreover, even the pessimistic parameter values that define Models A and B may be practically relevant, especially when the extent of OFF-source regions is restricted by nearby sources, or stringent event selection limits the overall count statistics. 

The joint likelihood method demonstrates good performance in a number of simple but experimentally relevant scenarios where the reliability of traditional data stacking is seriously impaired. Nonetheless, it should be stated that the \textit{full} potential of the joint likelihood method can only be realised if the distributions of all the systematic uncertainties affect each observational dataset are parameterisable with sufficient accuracy. 

Admittedly, definitive derivation of the distribution of $\alpha$ is unlikely to be straightforward in many situations. Semi-analytic estimation of the required distributions necessitates detailed knowledge of the instrumental response as well as careful measurement of the environmental conditions at the time of observation. Accordingly, imperfect or incomplete understanding of the dominant systematic biases might render this approach untenable. Plausible alternative strategies include jackknife or bootstrap resampling of OFF-source data from randomly selected background sampling regions to derive multiple estimates for $\alpha$ and construct an empirical distribution. In this case, the accuracy of the resultant parameterisation is limited by the finite number of permutations available for the resampling process. Emulating the approach of \citet{2011arXiv1112.0786K} and segregating data on the basis of quantifiable \textit{operating conditions} would enable the use of large representative datasets to derive template distributions that pertain to finite subsets of the observational parameter space. 

In situations where the suggested techniques are unable to accurately reconstruct the shape of the $\alpha$ distribution, a generic parameterisation could be tailored using target-specific estimates of the dispersion in derived values of the ON/OFF normalisation. By acknowledging the existence of the associated systematic uncertainties, this first-order approximation would ameliorate under-coverage as well as deviations from the nominal $\chi_{1}$ distribution for $S$, thereby reducing the rate of unexpected false-positive detections. Moreover, by facilitating limited adaptation of the modelled systematic uncertainties, this approach retains a key advantage of the full joint likelihood technique that cannot be straightforwardly emulated by traditional data stacking. Indeed, it is likely that this simplified approach might recover a significant proportion of the power which is achieved when the true distribution of $\alpha$ is available. Nonetheless, as demonstrated by the Monte Carlo results for a symmetric distribution of $\alpha$, there may be situations in which the advantages offered by the joint likelihood approach are marginal and the performance of the data stacking approach is deemed sufficient.

While this study has focussed upon parameterisation of $\alpha$ uncertainties, the joint likelihood technique represents a highly flexible and extensible analytical approach. Indeed, related analyses are able to parameterise arbitrary properties of the instrument response, the measured $\gamma$-ray signal, and even suspected source properties such as the spectral index and temporal variability \cite[e.g.][]{2011PhRvL.107x1302A}. 

\section{Conclusions}\label{sec:conclusion}

Toy Monte Carlo simulations have been used to investigate the effect of non-negligible systematic uncertainties affecting the relative normalisation of the ON and OFF-source event sampling regions in atmospheric Cherenkov telescope observations. As a specific example, the properties of two alternative strategies for combined source analysis have been examined. Situations have been identified in which the traditional \emph{data stacking} approach exhibits unexpected and undesirable statistical behaviour. 
Even when correctly calibrated null distributions for the combined significance are employed, the sensitivity of the data stacking approach appears markedly impaired if asymmetrically distributed $\alpha$ uncertainties are assumed. The data stacking technique is also yields unreliable estimates for the \gr\ signal parameters, with derived confidence intervals deviate from their stated coverage by a margin which widens with increasing target multiplicity.  

An alternative to data stacking has been outlined which combines target-specific likelihood functions that explicitly account for non-uniform systematic uncertainties. As an analytical approach, the \emph{joint likelihood} technique exhibits many characteristics which surpass those of traditional data stacking. It offers a statistically robust prescription for the treatment of target-specific systematic uncertainties, and effectively addresses the shortcomings which are inherent in the data stacking framework. Indeed, the joint likelihood effectively eliminates pathological behaviour inherent in data stacking, with combined source significances conforming to a single, nominal null distribution. Moreover, it achieves equivalent or superior sensitivity to the data stacking technique and yields parameter confidence intervals that exhibit minimal deviation from their stated coverage.

Many of the insights provided by this study of combined source analyses are equally applicable to  single-target analyses involving similarly abundant event statistics. Indeed, the synthesis of any extensive dataset often combines observations which incorporate diverse systematic effects. This study has shown that the joint likelihood method provides a viable approach for the robust combination of data with inhomogeneous systematic uncertainties, irrespective of the celestial target coordinates. Although it is a generally applicable technique, the joint likelihood analysis is likely to prove most useful in experimental situations involving limited event statistics and a restricted choice of OFF-source sampling regions.
  

\end{document}